\input harvmac
\input amssym
\baselineskip 13pt


\def\bigl{\Big(}
\def\bigr{\Big)}
\def\winf{$W_{\infty}[\l]$}
\def\p{\partial}
\def\Wc{\cal{W}}
\def\half{{1\over 2}}
\def\rar{\rightarrow}
\def\mw{{\cal{W}}}
\def\ml{{\cal{L}}}
\def\hsl{hs[$\lambda$]}
\def\vs{\vskip .1 in}
\def\a{\alpha}

\def\o{\omega}
\def\l{\lambda}
\def\t{\tau}

\def\ept{e^{2\rho}}

\def\p{\partial}
\def\bul{$\bullet$~}
\def\Ab{\overline{A}}

\def\slth{$sl(3,R)$}
\def\cg{{\cal{G}}}
\def\taub{\overline{\tau}}


\def\p{\partial}
\def\eps{\epsilon}
\def\epsb{\overline{\eps}}
\def\wb{\overline{w}}
\def\Lc{{\cal{L}}}
\def\Lcb{\overline{{\cal L}} }
\def\Ab{\overline{A}}
\def\zb{\overline{z}}
\def\ab{\overline{a}}


\def\Wc{{\cal{W}}}
\def\Wcb{\overline{{\cal W}}}
\def\mub{\bar{\mu}}


\lref\StromingerEQ{
  A.~Strominger,
  ``Black hole entropy from near-horizon microstates,''
  JHEP {\bf 9802}, 009 (1998)
  [arXiv:hep-th/9712251].
}

\lref\BalasubramanianRE{
  V.~Balasubramanian and P.~Kraus,
  ``A stress tensor for anti-de Sitter gravity,''
  Commun.\ Math.\ Phys.\  {\bf 208}, 413 (1999)
  [arXiv:hep-th/9902121].
}

\lref\fef{C. Fefferman and C.R. Graham, ``Conformal Invariants",
in {\it Elie Cartan et les Math\'{e}matiques d'aujourd'hui}
(Ast\'{e}risque, 1985) 95.}

\lref\deHaroXN{
  S.~de Haro, S.~N.~Solodukhin and K.~Skenderis,
  ``Holographic reconstruction of spacetime and renormalization in the  AdS/CFT
  correspondence,''
  Commun.\ Math.\ Phys.\  {\bf 217}, 595 (2001)
  [arXiv:hep-th/0002230].
}

\lref\PolchinskiRQ{
  J.~Polchinski,
  ``String theory, Vol. 1'', Cambridge University Press, 1998.
}

\lref\MaldacenaBW{
  J.~M.~Maldacena and A.~Strominger,
  ``AdS(3) black holes and a stringy exclusion principle,''
JHEP {\bf 9812}, 005 (1998).
[hep-th/9804085].
}

\lref\KrausVZ{
  P.~Kraus and F.~Larsen,
  ``Microscopic black hole entropy in theories with higher derivatives,''
JHEP {\bf 0509}, 034 (2005).
[hep-th/0506176].
}

\lref\KrausNB{
  P.~Kraus and F.~Larsen,
  ``Partition functions and elliptic genera from supergravity,''
JHEP {\bf 0701}, 002 (2007).
[hep-th/0607138].
}

\lref\AndradeSX{
  T.~Andrade, J.~I.~Jottar and R.~G.~Leigh,
  ``Boundary Conditions and Unitarity: the Maxwell-Chern-Simons System in AdS$_3$/CFT$_2$,''
JHEP {\bf 1205}, 071 (2012).
[arXiv:1111.5054 [hep-th]].
}

\lref\SezginPV{
  E.~Sezgin and P.~Sundell,
  ``An Exact solution of 4-D higher-spin gauge theory,''
Nucl.\ Phys.\ B {\bf 762}, 1 (2007).
[hep-th/0508158].
}

\lref\IazeollaWT{
  C.~Iazeolla, E.~Sezgin and P.~Sundell,
  ``Real forms of complex higher spin field equations and new exact solutions,''
Nucl.\ Phys.\ B {\bf 791}, 231 (2008).
[arXiv:0706.2983 [hep-th]].
}

\lref\DidenkoVA{
  V.~E.~Didenko, A.~S.~Matveev and M.~A.~Vasiliev,
  ``Unfolded Description of AdS(4) Kerr Black Hole,''
Phys.\ Lett.\ B {\bf 665}, 284 (2008).
[arXiv:0801.2213 [gr-qc]].
}

\lref\AhnBY{
  C.~Ahn,
  ``The Coset Spin-4 Casimir Operator and Its Three-Point Functions with Scalars,''
JHEP {\bf 1202}, 027 (2012).
[arXiv:1111.0091 [hep-th]].
}

\lref\DidenkoTD{
  V.~E.~Didenko and M.~A.~Vasiliev,
  ``Static BPS black hole in 4d higher-spin gauge theory,''
Phys.\ Lett.\ B {\bf 682}, 305 (2009).
[arXiv:0906.3898 [hep-th]].
}

\lref\SezginRU{
  E.~Sezgin and P.~Sundell,
  ``Analysis of higher spin field equations in four-dimensions,''
JHEP {\bf 0207}, 055 (2002).
[hep-th/0205132].
}

\lref\KrausWN{
  P.~Kraus,
  ``Lectures on black holes and the AdS(3)/CFT(2) correspondence,''
  Lect.\ Notes Phys.\  {\bf 755}, 193 (2008)
  [arXiv:hep-th/0609074].
}

\lref\BanadosGG{
  M.~Banados,
  ``Three-dimensional quantum geometry and black holes,''
  arXiv:hep-th/9901148.
}

\lref\IazeollaCB{
  C.~Iazeolla and P.~Sundell,
  ``Families of exact solutions to Vasiliev's 4D equations with spherical, cylindrical and biaxial symmetry,''
JHEP {\bf 1112}, 084 (2011).
[arXiv:1107.1217 [hep-th]].
}

\lref\ShenkerZF{
  S.~H.~Shenker and X.~Yin,
  ``Vector Models in the Singlet Sector at Finite Temperature,''
[arXiv:1109.3519 [hep-th]].
}

\lref\MaldacenaRE{
  J.~M.~Maldacena,
  ``The Large N limit of superconformal field theories and supergravity,''
Adv.\ Theor.\ Math.\ Phys.\  {\bf 2}, 231 (1998).
[hep-th/9711200].
}

\lref\GubserBC{
  S.~S.~Gubser, I.~R.~Klebanov and A.~M.~Polyakov,
  ``Gauge theory correlators from noncritical string theory,''
Phys.\ Lett.\ B {\bf 428}, 105 (1998).
[hep-th/9802109].
}

\lref\WittenQJ{
  E.~Witten,
  ``Anti-de Sitter space and holography,''
Adv.\ Theor.\ Math.\ Phys.\  {\bf 2}, 253 (1998).
[hep-th/9802150].
}

\lref\StromingerSH{
  A.~Strominger and C.~Vafa,
  ``Microscopic origin of the Bekenstein-Hawking entropy,''
Phys.\ Lett.\ B {\bf 379}, 99 (1996).
[hep-th/9601029].
}

\lref\BrownNW{
  J.~D.~Brown and M.~Henneaux,
  ``Central Charges in the Canonical Realization of Asymptotic Symmetries: An Example from Three-Dimensional Gravity,''
Commun.\ Math.\ Phys.\  {\bf 104}, 207 (1986)..
}

\lref\BanadosGQ{
  M.~Banados, M.~Henneaux, C.~Teitelboim and J.~Zanelli,
  ``Geometry of the (2+1) black hole,''
Phys.\ Rev.\ D {\bf 48}, 1506 (1993).
[gr-qc/9302012].
}

\lref\BanadosWN{
  M.~Banados, C.~Teitelboim and J.~Zanelli,
  ``The Black hole in three-dimensional ,''
Phys.\ Rev.\ Lett.\  {\bf 69}, 1849 (1992).
[hep-th/9204099].
}

\lref\FronsdalRB{
  C.~Fronsdal,
  ``Massless Fields with Integer Spin,''
Phys.\ Rev.\ D {\bf 18}, 3624 (1978)..
}

\lref\FradkinKS{
  E.~S.~Fradkin and M.~A.~Vasiliev,
  ``On the Gravitational Interaction of Massless Higher Spin Fields,''
Phys.\ Lett.\ B {\bf 189}, 89 (1987)..
}

\lref\FradkinQY{
  E.~S.~Fradkin and M.~A.~Vasiliev,
  ``Cubic Interaction in Extended Theories of Massless Higher Spin Fields,''
Nucl.\ Phys.\ B {\bf 291}, 141 (1987)..
}

\lref\VasilievEN{
  M.~A.~Vasiliev,
  ``Consistent equation for interacting gauge fields of all spins in (3+1)-dimensions,''
Phys.\ Lett.\ B {\bf 243}, 378 (1990)..
}

\lref\VasilievAV{
  M.~A.~Vasiliev,
  ``More on equations of motion for interacting massless fields of all spins in (3+1)-dimensions,''
Phys.\ Lett.\ B {\bf 285}, 225 (1992)..
}

\lref\VasilievRN{
  M.~A.~Vasiliev,
  ``Higher spin symmetries, star product and relativistic equations in AdS space,''
  hep-th/0002183.
  }

\lref\KlebanovJA{
  I.~R.~Klebanov and A.~M.~Polyakov,
  ``AdS dual of the critical O(N) vector model,''
Phys.\ Lett.\ B {\bf 550}, 213 (2002).
[hep-th/0210114].
}

\lref\ProkushkinBQ{
  S.~F.~Prokushkin and M.~A.~Vasiliev,
  ``Higher spin gauge interactions for massive matter fields in 3-D AdS space-time,''
Nucl.\ Phys.\ B {\bf 545}, 385 (1999).
[hep-th/9806236].
}

\lref\WittenHC{
  E.~Witten,
  ``(2+1)-Dimensional Gravity as an Exactly Soluble System,''
  Nucl.\ Phys.\  B {\bf 311}, 46 (1988).
}

\lref\AchucarroVZ{
  A.~Achucarro and P.~K.~Townsend,
  ``A Chern-Simons Action for Three-Dimensional anti-De Sitter Supergravity
  Theories,''
  Phys.\ Lett.\  B {\bf 180}, 89 (1986).
}

\lref\PetkouZZ{
  A.~C.~Petkou,
  ``Evaluating the AdS dual of the critical O(N) vector model,''
JHEP {\bf 0303}, 049 (2003).
[hep-th/0302063].
}

\lref\LeighGK{
  R.~G.~Leigh and A.~C.~Petkou,
  ``Holography of the N=1 higher spin theory on AdS(4),''
JHEP {\bf 0306}, 011 (2003).
[hep-th/0304217].
}

\lref\CastroBC{
  A.~Castro, E.~Hijano and A.~Lepage-Jutier,
  ``Unitarity Bounds in AdS$_3$ Higher Spin Gravity,''
JHEP {\bf 1206}, 001 (2012).
[arXiv:1202.4467 [hep-th]].
}

\lref\zhu{
Y.~Zhu,
``Modular invariance of characters of vertex operator algebras,"
 J.\ Amer.\ Math.\ Soc.\ {\bf 9}, 237 (1996).}

\lref\SezginPT{
  E.~Sezgin and P.~Sundell,
  ``Holography in 4D (super) higher spin theories and a test via cubic scalar couplings,''
JHEP {\bf 0507}, 044 (2005).
[hep-th/0305040].
}

\lref\TanTJ{
  H.~-S.~Tan,
  ``Aspects of Three-dimensional Spin-4 Gravity,''
JHEP {\bf 1202}, 035 (2012).
[arXiv:1111.2834 [hep-th]].
}

\lref\GiombiYA{
  S.~Giombi and X.~Yin,
  ``On Higher Spin Gauge Theory and the Critical O(N) Model,''
[arXiv:1105.4011 [hep-th]].
}

\lref\GiombiWH{
  S.~Giombi and X.~Yin,
  ``Higher Spin Gauge Theory and Holography: The Three-Point Functions,''
JHEP {\bf 1009}, 115 (2010).
[arXiv:0912.3462 [hep-th]].
}

\lref\GiombiVG{
  S.~Giombi and X.~Yin,
  ``Higher Spins in AdS and Twistorial Holography,''
JHEP {\bf 1104}, 086 (2011).
[arXiv:1004.3736 [hep-th]].
}

\lref\KochCY{
  R.~d.~M.~Koch, A.~Jevicki, K.~Jin and J.~P.~Rodrigues,
  ``$AdS_4/CFT_3$ Construction from Collective Fields,''
Phys.\ Rev.\ D {\bf 83}, 025006 (2011).
[arXiv:1008.0633 [hep-th]].
}

\lref\DouglasRC{
  M.~R.~Douglas, L.~Mazzucato and S.~S.~Razamat,
  ``Holographic dual of free field theory,''
Phys.\ Rev.\ D {\bf 83}, 071701 (2011).
[arXiv:1011.4926 [hep-th]].
}

\lref\MikhailovBP{
  A.~Mikhailov,
  ``Notes on higher spin symmetries,''
[hep-th/0201019].
}

\lref\DavidXG{
  J.~R.~David, M.~RGaberdiel and R.~Gopakumar,
  ``The Heat Kernel on AdS(3) and its Applications,''
JHEP {\bf 1004}, 125 (2010).
[arXiv:0911.5085 [hep-th]].
}

\lref\BekaertHW{
  X.~Bekaert, N.~Boulanger and P.~Sundell,
  ``How higher-spin gravity surpasses the spin two barrier: no-go theorems versus yes-go examples,''
[arXiv:1007.0435 [hep-th]].
}

\lref\GaberdielAR{
  M.~R.~Gaberdiel, R.~Gopakumar and A.~Saha,
  ``Quantum $W$-symmetry in $AdS_3$,''
JHEP {\bf 1102}, 004 (2011).
[arXiv:1009.6087 [hep-th]].
}

\lref\GaberdielPZ{
  M.~R.~Gaberdiel and R.~Gopakumar,
  ``An $AdS_3$ Dual for Minimal Model CFTs,''
Phys.\ Rev.\ D {\bf 83}, 066007 (2011).
[arXiv:1011.2986 [hep-th]].
}

\lref\GaberdielZW{
  M.~R.~Gaberdiel, R.~Gopakumar, T.~Hartman and S.~Raju,
  ``Partition Functions of Holographic Minimal Models,''
JHEP {\bf 1108}, 077 (2011).
[arXiv:1106.1897 [hep-th]].
}

\lref\GaberdielKU{
  M.~R.~Gaberdiel and R.~Gopakumar,
  ``Triality in Minimal Model Holography,''
[arXiv:1205.2472 [hep-th]].
}

\lref\BlencoweGJ{
  M.~P.~Blencowe,
  ``A Consistent Interacting Massless Higher Spin Field Theory In D = (2+1),''
  Class.\ Quant.\ Grav.\  {\bf 6}, 443 (1989).
}

\lref\BergshoeffNS{
  E.~Bergshoeff, M.~P.~Blencowe and K.~S.~Stelle,
  ``Area Preserving Diffeomorphisms And Higher Spin Algebra,''
  Commun.\ Math.\ Phys.\  {\bf 128}, 213 (1990).
}

\lref\BershadskyBG{
  M.~Bershadsky,
  ``Conformal field theories via Hamiltonian reduction,''
  Commun.\ Math.\ Phys.\  {\bf 139}, 71 (1991).
}

\lref\PolyakovDM{
  A.~M.~Polyakov,
  ``Gauge Transformations and Diffeomorphisms,''
  Int.\ J.\ Mod.\ Phys.\  A {\bf 5}, 833 (1990).
}

\lref\BilalCF{
  A.~Bilal,
  ``W algebras from Chern-Simons theory,''
  Phys.\ Lett.\  B {\bf 267}, 487 (1991).
}

\lref\GaberdielWB{
  M.~R.~Gaberdiel and T.~Hartman,
  ``Symmetries of Holographic Minimal Models,''
  arXiv:1101.2910 [hep-th].
}

\lref\HenneauxXG{
  M.~Henneaux and S.~J.~Rey,
  ``Nonlinear W(infinity) Algebra as Asymptotic Symmetry of Three-Dimensional
  Higher Spin Anti-de Sitter Gravity,''
  JHEP {\bf 1012}, 007 (2010)
  [arXiv:1008.4579 [hep-th]].
}

\lref\CampoleoniZQ{
  A.~Campoleoni, S.~Fredenhagen, S.~Pfenninger and S.~Theisen,
  ``Asymptotic symmetries of three-dimensional gravity coupled to higher-spin
  fields,''
  JHEP {\bf 1011}, 007 (2010)
  [arXiv:1008.4744 [hep-th]].
}

\lref\CampoleoniHG{
  A.~Campoleoni, S.~Fredenhagen and S.~Pfenninger,
  ``Asymptotic W-symmetries in three-dimensional higher-spin gauge theories,''
JHEP {\bf 1109}, 113 (2011).
[arXiv:1107.0290 [hep-th]].
}

\lref\AmmonUA{
  M.~Ammon, P.~Kraus and E.~Perlmutter,
  ``Scalar fields and three-point functions in D=3 higher spin gravity,''
[arXiv:1111.3926 [hep-th]].
}

\lref\ChangMZ{
  C.~-M.~Chang and X.~Yin,
  ``Higher Spin Gravity with Matter in $AdS_3$ and Its CFT Dual,''
[arXiv:1106.2580 [hep-th]].
}

\lref\MaldacenaSF{
  J.~Maldacena and A.~Zhiboedov,
  ``Constraining conformal field theories with a slightly broken higher spin symmetry,''
[arXiv:1204.3882 [hep-th]].
}

\lref\MaldacenaJN{
  J.~Maldacena and A.~Zhiboedov,
  ``Constraining Conformal Field Theories with A Higher Spin Symmetry,''
[arXiv:1112.1016 [hep-th]].
}

\lref\GiombiKC{
  S.~Giombi, S.~Minwalla, S.~Prakash, S.~P.~Trivedi, S.~R.~Wadia and X.~Yin,
  ``Chern-Simons Theory with Vector Fermion Matter,''
[arXiv:1110.4386 [hep-th]].
}

\lref\AharonyJZ{
  O.~Aharony, G.~Gur-Ari and R.~Yacoby,
  ``d=3 Bosonic Vector Models Coupled to Chern-Simons Gauge Theories,''
JHEP {\bf 1203}, 037 (2012).
[arXiv:1110.4382 [hep-th]].
}

\lref\BekaertUX{
  X.~Bekaert, E.~Joung and J.~Mourad,
  ``Comments on higher-spin holography,''
[arXiv:1202.0543 [hep-th]].
}

\lref\CanduJQ{
  C.~Candu and M.~R.~Gaberdiel,
  ``Supersymmetric holography on $AdS_3$,''
[arXiv:1203.1939 [hep-th]].
}

\lref\HenneauxNY{
  M.~Henneaux, G.~Lucena Gomez, J.~Park and S.~-J.~Rey,
  ``Super- W(infinity) Asymptotic Symmetry of Higher-Spin $AdS_3$ Supergravity,''
JHEP {\bf 1206}, 037 (2012).
[arXiv:1203.5152 [hep-th]].
}

\lref\VasilievBA{
  M.~A.~Vasiliev,
  ``Higher spin gauge theories: Star product and AdS space,''
In *Shifman, M.A. (ed.): The many faces of the superworld* 533-610.
[hep-th/9910096].
}

\lref\AhnPV{
  C.~Ahn,
  ``The Large N 't Hooft Limit of Coset Minimal Models,''
JHEP {\bf 1110}, 125 (2011).
[arXiv:1106.0351 [hep-th]].
}

\lref\GaberdielNT{
  M.~RGaberdiel and C.~Vollenweider,
  ``Minimal Model Holography for SO(2N),''
JHEP {\bf 1108}, 104 (2011).
[arXiv:1106.2634 [hep-th]].
}

\lref\AhnFZ{
  C.~Ahn,
  ``The Large N 't Hooft Limit of Kazama-Suzuki Model,''
[arXiv:1206.0054 [hep-th]].
}

\lref\PapadodimasPF{
  K.~Papadodimas and S.~Raju,
  ``Correlation Functions in Holographic Minimal Models,''
Nucl.\ Phys.\ B {\bf 856}, 607 (2012).
[arXiv:1108.3077 [hep-th]].
}

\lref\CastroIW{
  A.~Castro, R.~Gopakumar, M.~Gutperle and J.~Raeymaekers,
  ``Conical Defects in Higher Spin Theories,''
JHEP {\bf 1202}, 096 (2012).
[arXiv:1111.3381 [hep-th]].
}

\lref\HanakiYF{
  K.~Hanaki and C.~Peng,
  ``Symmetries of Holographic Super-Minimal Models,''
[arXiv:1203.5768 [hep-th]].
}

\lref\GutperleKF{
  M.~Gutperle and P.~Kraus,
  ``Higher Spin Black Holes,''
JHEP {\bf 1105}, 022 (2011).
[arXiv:1103.4304 [hep-th]].
}

\lref\AmmonNK{
  M.~Ammon, M.~Gutperle, P.~Kraus and E.~Perlmutter,
  ``Spacetime Geometry in Higher Spin Gravity,''
JHEP {\bf 1110}, 053 (2011).
[arXiv:1106.4788 [hep-th]].
}

\lref\KrausDS{
  P.~Kraus and E.~Perlmutter,
  ``Partition functions of higher spin black holes and their CFT duals,''
JHEP {\bf 1111}, 061 (2011).
[arXiv:1108.2567 [hep-th]].
}

\lref\GaberdielYB{
  M.~R.~Gaberdiel, T.~Hartman and K.~Jin,
  ``Higher Spin Black Holes from CFT,''
JHEP {\bf 1204}, 103 (2012).
[arXiv:1203.0015 [hep-th]].
}

\lref\BanadosUE{
  M.~Banados, R.~Canto and S.~Theisen,
  ``The Action for higher spin black holes in three dimensions,''
[arXiv:1204.5105 [hep-th]].
}

\lref\HawkingSW{
  S.~W.~Hawking,
  ``Particle Creation by Black Holes,''
Commun.\ Math.\ Phys.\  {\bf 43}, 199 (1975), [Erratum-ibid.\  {\bf 46}, 206 (1976)]..
}

\lref\MaldacenaKR{
  J.~M.~Maldacena,
  ``Eternal black holes in anti-de Sitter,''
JHEP {\bf 0304}, 021 (2003).
[hep-th/0106112].
}

\lref\KrausIV{
  P.~Kraus, H.~Ooguri and S.~Shenker,
  ``Inside the horizon with AdS / CFT,''
Phys.\ Rev.\ D {\bf 67}, 124022 (2003).
[hep-th/0212277].
}

\lref\FidkowskiNF{
  L.~Fidkowski, V.~Hubeny, M.~Kleban and S.~Shenker,
  ``The Black hole singularity in AdS / CFT,''
JHEP {\bf 0402}, 014 (2004).
[hep-th/0306170].
}

\lref\SundborgWP{
  B.~Sundborg,
  ``Stringy gravity, interacting tensionless strings and massless higher spins,''
Nucl.\ Phys.\ Proc.\ Suppl.\  {\bf 102}, 113 (2001).
[hep-th/0103247].
}

\lref\PopeSR{
  C.~N.~Pope, L.~J.~Romans and X.~Shen,
  ``W(infinity) And The Racah-wigner Algebra,''
Nucl.\ Phys.\ B {\bf 339}, 191 (1990)..
}

\lref\CastroFM{
  A.~Castro, E.~Hijano, A.~Lepage-Jutier and A.~Maloney,
  ``Black Holes and Singularity Resolution in Higher Spin Gravity,''
JHEP {\bf 1201}, 031 (2012).
[arXiv:1110.4117 [hep-th]].
}

\lref\WaldNT{
  R.~M.~Wald,
  ``Black hole entropy is the Noether charge,''
  Phys.\ Rev.\  D {\bf 48}, 3427 (1993)
  [arXiv:gr-qc/9307038].
}

\lref\CreutzigFE{
  T.~Creutzig, Y.~Hikida and P.~B.~Ronne,
  ``Higher spin $AdS_3$ supergravity and its dual CFT,''
JHEP {\bf 1202}, 109 (2012).
[arXiv:1111.2139 [hep-th]].
}

\lref\GirardelloPP{
  L.~Girardello, M.~Porrati and A.~Zaffaroni,
  ``3-D interacting CFTs and generalized Higgs phenomenon in higher spin theories on AdS,''
Phys.\ Lett.\ B {\bf 561}, 289 (2003).
[hep-th/0212181].
}

\lref\PE{P. Kraus, E. Perlmutter, in progress.}

\lref\GibbonsUE{
 G.~W.~Gibbons and S.~W.~Hawking,
 ``Action Integrals and Partition Functions in Quantum Gravity,''
Phys.\ Rev.\ D {\bf 15}, 2752 (1977)..
}

\lref\ZamolodchikovWN{
  A.~B.~Zamolodchikov,
  ``Infinite Additional Symmetries In Two-Dimensional Conformal Quantum Field
  Theory,''
  Theor.\ Math.\ Phys.\  {\bf 65}, 1205 (1985)
  [Teor.\ Mat.\ Fiz.\  {\bf 65}, 347 (1985)].
}

\lref\IazeollaNF{
  C.~Iazeolla and P.~Sundell,
  ``Biaxially symmetric solutions to 4D higher-spin gravity,''
[arXiv:1208.4077 [hep-th]].
}

\lref\DattaKM{
  S.~Datta and J.~R.~David,
  ``Supersymmetry of classical solutions in Chern-Simons higher spin supergravity,''
[arXiv:1208.3921 [hep-th]].
}

\lref\CanduTR{
  C.~Candu and M.~R.~Gaberdiel,
  ``Duality in N=2 minimal model holography,''
[arXiv:1207.6646 [hep-th]].
}

\lref\TanXI{
  H.~S.~Tan,
  ``Exploring Three-dimensional Higher-Spin Supergravity based on sl(N |N - 1) Chern-Simons theories,''
[arXiv:1208.2277 [hep-th]].
}

\lref\CampoleoniHP{
  A.~Campoleoni, S.~Fredenhagen, S.~Pfenninger and S.~Theisen,
  ``Towards metric-like higher-spin gauge theories in three dimensions,''
[arXiv:1208.1851 [hep-th]].
}

\lref\PerezCF{
  A.~Perez, D.~Tempo and R.~Troncoso,
  ``Higher spin gravity in 3D: black holes, global charges and thermodynamics,''
[arXiv:1207.2844 [hep-th]].
}

\Title{\vbox{\baselineskip14pt
}} {\vbox{\centerline {Black holes in three dimensional}
\medskip\vbox{\centerline {higher spin gravity: A review}}}}
\centerline{Martin Ammon, Michael Gutperle, Per
Kraus and Eric Perlmutter\foot{ammon@physics.ucla.edu, gutperle@ucla.edu, pkraus@ucla.edu, perl@physics.ucla.edu}}
\bigskip
\centerline{\it{Department of Physics and Astronomy}}
\centerline{${}$\it{University of California, Los Angeles, CA 90095, USA}}

\baselineskip14pt

\vskip .3in

\centerline{\bf Abstract}
\vskip.2cm
We review recent progress in the construction of black holes in three dimensional higher spin gravity theories. Starting from spin-3 gravity and working our way toward the theory of an infinite tower of higher spins coupled to matter, we show how to harness higher spin gauge invariance to consistently generalize familiar notions of black holes. We review the construction of black holes with conserved higher spin charges and the computation of their partition functions to leading asymptotic order. In view of the AdS/CFT correspondence as applied to certain vector-like conformal field theories with extended conformal symmetry, we successfully compare to CFT calculations in a generalized Cardy regime. A brief recollection of pertinent aspects of ordinary gravity is also given.

\Date{July  2012}
\baselineskip13pt

\listtoc 

\writetoc

\newsec{Introduction}

Black holes play  a central role in both classical and quantum gravitational physics.   In classical general relativity black holes constitute an important class of exact solutions of Einstein's equations, where questions such as the causal structure of spacetime and  the nature of singularities can be addressed. There is overwhelming evidence that black holes exist as astrophysical objects in binary systems and at the center of galaxies.

The discovery of  Hawking \HawkingSW\  that black holes radiate as black bodies leads to the black hole information paradox, whereby in the process of black hole formation and evaporation pure states seem to evolve into mixed states and hence information (and consequently unitary time evolution in quantum mechanics) is not preserved.
Any theory of quantum gravity has to resolve this clash between classical gravity and quantum mechanics.  String theory has made important steps in this direction. First, string theory  provides a microscopic accounting of the Bekenstein-Hawking entropy  of certain near extremal black holes found in superstring theories \StromingerSH. Second, the AdS/CFT correspondence \refs{\MaldacenaRE,\GubserBC,\WittenQJ} provides a conceptual  resolution of the information paradox for black holes in asymptotically anti-de Sitter spacetimes.  As time evolution on the CFT side is unitary,  the time evolution on the dual bulk gravity side must be unitary too.
These arguments  give credence to the point of view that the information of an evaporating black hole is somehow encoded in the outgoing Hawking radiation.

The entropy counting in string theory generically involves a continuation from weak to strong coupling and works most reliably for   black holes  in theories which enjoy non-renormalization due to  a large amount  of supersymmetry. One needs  to go beyond  supersymmetric examples in order to obtain a microscopic counting of   the entropy for Schwarzschild black holes. Unfortunately this goal has not  yet been achieved.  It is therefore useful to consider black holes in  simplified theories  and toy models, where supersymmetry and non renormalization theorems are not necessary to perform calculations reliably.

One such example is three dimensional gravity. The main simplifying feature of this theory is that a graviton  in three dimensions does not have any propagating
degrees of freedom.  However, in  asymptotically AdS space the theory has nontrivial dynamics given by the dual two-dimensional CFT which is localized at the boundary of AdS$_3$.   A reflection of the topological nature of three
dimensional  gravity  is the fact that  with a negative cosmological constant  Einstein gravity can be reformulated as a Chern-Simons gauge theory
with gauge algebra $sl(2,R)\oplus sl(2,R)$ \refs{\WittenHC,\AchucarroVZ}. The diffeomorphism and local Lorentz symmetries of the gravitational theory  are reinterpreted as gauge transformations of the Chern-Simons theory.  In
\BrownNW\  it was discovered that the generators of the  diffeormorphism  transformations that preserve  the asymptotic form of the  AdS metric form the Virasoro algebra, to be viewed as the  algebra of  conformal
symmetries  in the dual CFT. In the Chern-Simons formulation of the theory the Virasoro generators manifest themselves in terms of Hamiltonian reduction of the gauge connection.

In three dimensional gravity with a negative cosmological constant  BTZ black hole solutions  \refs{\BanadosWN,\BanadosGQ} can be constructed as orbifolds of the AdS$_3$ vacuum. These   black hole solutions are
 simpler than their higher dimensional cousins as they are locally still given by the AdS$_3$ vacuum.  The BTZ black hole  nevertheless has a horizon, finite entropy and nonzero Hawking temperature. In the limit of high temperature the entropy of the black hole is related to the central charge of the CFT via the Cardy formula. The simplicity of the BTZ black hole allows one to address difficult  conceptual questions
related to black holes. Two examples are  the nature of entanglement of degrees of freedom inside and outside the horizon \MaldacenaKR\ and whether it is possible to access  the region behind the black hole  horizon by correlation functions of operators localized on the boundary  \refs{\KrausIV,\FidkowskiNF}.

In the AdS/CFT correspondence bulk calculations can be most easily performed in the weak coupling, small curvature regime   where semi-classical gravity methods are sufficient. However, in this limit we lose several characteristic features of string theory such as the non-locality due to the finite extent of strings, the infinite tower of fields of ever increasing spin and the generalization of diffeomorphism invariance of gravity to a very large gauge symmetry which is  manifest in string field theory.

An interesting theory  has been developed over the last 20 years by Vasiliev and collaborators which   is in some sense  located ``in-between" Einstein gravity and full fledged string theory. It is an extension of ordinary gravity with a negative cosmological constant coupled to an infinite tower of massless fields of higher
spin in three or more dimensions. The classical equations of motion are explicitly known, and are nonlocal as well as highly
nonlinear. Note however that at present no action is known which reproduces the equations of motion.  The no-go theorems in flat space are evaded due to the absence of an S-matrix in AdS space; see, e.g., \BekaertHW\ for a modern discussion.
The higher spin theory has a  large higher spin gauge symmetry which generalizes the spin 2 gauge symmetry of ordinary gravity. A partial list of the original papers is     \refs{\FronsdalRB,\FradkinKS,\FradkinQY,\VasilievEN,\VasilievAV}, see also the reviews \refs{\VasilievRN,\VasilievBA}, which contain  more comprehensive lists of references.
  There exists some speculation that the higher spin   theories might be related to
    a tensionless limit of (topological) string theory \SundborgWP. These questions as well as the question of whether  classical Vasiliev theory theory can be gauge fixed and quantized while preserving the higher spin symmetries remain open at present.

There has been a renewed interest in Vasiliev theory due to the  duality conjecture that  relates the large $N$ limit of vector models in three dimensions  to higher spin gravity theories in AdS$_4$. The most  developed form of this
 conjecture is due to Klebanov and Polyakov \KlebanovJA, stating that  Vasiliev theory in AdS$_4$ is dual to the large $N$ limit of the three dimensional $O(N)$ vector model.  The massless higher spin fields in the bulk holographically realize the  infinite set of higher spin conserved  currents in the
 (free) UV boundary theory.  An impressive amount of evidence for the duality has been obtained in recent years  \refs{\MikhailovBP,\PetkouZZ,\LeighGK,\SezginPT,
 \KochCY,\DouglasRC,\GirardelloPP}, including the calculation of three point functions in the
 bulk \refs{\GiombiWH,\GiombiVG,\GiombiYA}. A proof of the duality relation has recently been obtained with   some assumptions which are very reasonable from the point of view of AdS/CFT \refs{\MaldacenaSF,\MaldacenaJN}. Other examples of dualities between higher spin theories  in $AdS_4$ and vector like theories in three dimensions were  presented in \refs{\GiombiKC,\AharonyJZ,\BekaertUX}.

In the present review we will focus on black holes in higher spin gravity in three dimensions.   A significant simplification over higher spin theories in four or more dimensions is the fact that the theory can be implemented
as a Chern-Simons gauge theory \refs{\BlencoweGJ,\BergshoeffNS,\CampoleoniZQ}. In its simplest realization the $sl(2,R)\oplus  sl(2,R)$ Chern Simons gauge group is replaced by $sl(N,R)\oplus  sl(N,R)$. Ordinary gravity is contained as a subsector of the theory and realized  by the principal embedding of  $sl(2,R)$ inside $sl(N,R)$.  The theory describes
massless interacting fields of spin $s=2,3,\cdots,N$.  The simplest case  $N=3$ describes ordinary gravity coupled to a massless spin three field \refs{\CampoleoniZQ, \CampoleoniHG} and exemplifies a general feature of the higher spin gravity theories: the gauge transformations include coordinate transformations as well as well as spin 3
gauge transformations under which the metric $g_{\mu\nu}$ transforms in a nontrivial way.  Consequently, diffeomorphism invariant quantities   such as the Ricci scalar are not gauge invariant under the higher spin gauge transformations.
Similarly,  global aspects of the spacetime such as its causal structure and the existence of a horizon are also not gauge invariant in  higher spin gravity.

The three dimensional Vasiliev theory is a one parameter family of theories based on an infinite-dimensional gauge algebra \hsl\ \PopeSR. For these theories it is possible to consistently couple propagating matter fields to the infinite set of higher spin fields.  Note that there appears to be little understanding of how to consistently couple matter to higher spin theories for finite dimensional gauge algebras such as $sl(N,R)$.

The asymptotic symmetries of the higher spin gravity can most easily be obtained by Hamiltonian reduction \refs{ \PolyakovDM,\BershadskyBG,\BilalCF,\CampoleoniZQ,\HenneauxXG,\GaberdielWB}. This construction  generalizes  the realization of the Virasoro algebra for the $sl(2,R)$ case of pure gravity. Instead of the Virasoro algebra one obtains $W_N$ algebras as the asymptotic symmetry of the dual CFTs.  Note that just like pure gravity the higher spin gravity theories based on $sl(N,R)$ have no propagating bulk degrees of freedom. Similarly for the \hsl\ theory, the asymptotic symmetry is a particular $W$-algebra known as \winf.

A concrete connection between bulk and boundary is given by a conjectured AdS/CFT duality due to Gaberdiel and Gopakumar \refs{\GaberdielPZ, \GaberdielKU}.  This conjecture  relates the three dimensional Vasiliev theory to a large $N$ limit of $W_N$ minimal models.  The  $W_N$ minimal model CFT has a representation in terms of a coset theory   and the 't Hooft like limit is defined as follows:

\eqn\coseta{
{su(N)_k\oplus  su(N)_1\over su(N)_{k+1}}, \quad k,N\to \infty, \quad \lambda={N\over k+N}\;\; {\rm fixed}}
where the 't Hooft coupling $\lambda$ is identified with the deformation parameter of the higher spin algebra  \hsl. The central charge of the CFT  is given in the large $N$ limit by
\eqn\centrala{
c\sim N(1-\lambda^2)~.}

The fact that the central charge grows like $N$ instead of $N^2$ makes this duality analogous to the  Klebanov-Polyakov   dual of vector models in AdS$_4$.   A very appealing feature of this example of a AdS/CFT duality is that the  coset dual CFT is very well under control. In addition both the bulk and boundary have a huge symmetry group which strongly constrains the dynamics.

A considerable amount of evidence for this conjecture has been obtained recently.
For example the symmetries and the spectrum of the bulk and boundary CFT have been matched in \refs{\GaberdielAR,\GaberdielKU,\GaberdielWB}. The calculation of the 1-loop contribution to the bulk partition function utilizing heat-kernel methods  \refs{ \DavidXG,\GaberdielZW}
agrees with the large N limit of the CFT partition function of the coset theories \coseta. In \refs{ \ChangMZ,\PapadodimasPF,\AmmonUA, \AhnBY}   certain  classes  of correlation functions were  calculated both in the bulk and boundary theory and were found
to match.  Some open questions remain, such as the exact structure of the symmetry algebra of the hs[$\l$] theory at the quantum level, i.e. for finite central charge as well as  the identification of certain  light states which are present in the CFT with bulk configurations.  For recent progress towards resolving these questions, and a slight modification of the original conjecture of \GaberdielPZ, see \refs{\GaberdielKU,\CastroIW}.

In \refs{\GaberdielNT,\AhnPV} the Gaberdiel-Gopakumar conjecture has been generalized to other coset CFTs such as $so(2N)_k \oplus so(2N)_1/so(2n)_{k+1}$ coset, which is conjectured to be dual to a Vasiliev theory with massless fields with even spin  only.  Furthermore versions of the conjecture which relate superconformal cosets and supersymmetric higher spin theories were put forward  in \refs{ \CanduJQ,\AhnFZ,\HenneauxNY,\HanakiYF,\CreutzigFE,\CanduTR,\TanXI,\DattaKM}.

Black holes are important laboratories   to explore both perturbative and non perturbative aspects of this duality further.
An immediate conceptual challenge arises due to the existence of the higher spin gauge symmetries. In ordinary gravity the global causal structure of a spacetime, in particular the existence of a black hole horizon and singularities are diffeomorphism invariant properties of the spacetime. In higher spin gravity however these properties are not invariant under higher spin gauge transformations and consequently it is not at all obvious how we define  a black hole  in higher spin gravity.

Since it is possible to canonically embed $sl(2,R)$ as a sub algebra into both $sl(N,R)$ and \hsl,   a solution of ordinary gravity can always be embedded into a solution of  higher spin gravity.  Consequently, the BTZ black hole solution of ordinary three dimensional gravity is always a solution of higher spin gravity.
However  the presence of higher spin fields raises the question of how to generalize the BTZ black hole to a black hole which carries conserved higher spin charges.

 We would like  to find a gauge invariant characterization of what constitutes a  black hole in the higher spin gravity. In \GutperleKF\ a criterion for what constitutes a black hole was presented,  which states that  the   holonomies of the gauge connections around the contractible Euclidean time cycle  have to be  identical  to the holonomies of the gauge connection of the  embedded  BTZ black hole. This condition leads to  sensible  black hole  thermodynamics, i.e. the black hole has a  free energy which satisfies integrability conditions that are equivalent  to the first law of thermodynamics.   In  \refs{\AmmonNK,\CastroFM} it was shown that if the holonomy conditions are satisfied, there exists a higher spin gauge transformation which makes the solution manifestly a black hole, i.e. there is a horizon and the higher spin fields are smooth on the horizon.
On the dual CFT side the solution describes the the finite temperature partition function with a nonzero chemical potential for the higher spin $W$ charge. Consequently, the  holographic calculation of the partition functions  produces nontrivial predictions on the CFT side  \refs{ \KrausDS,\BanadosUE}, which can be checked by direct CFT calculations \GaberdielYB\ when a CFT with the right symmetries and large central charge is available.

The structure of this review is as follows. In section 2 we give a brief review of ordinary three dimensional gravity, its Chern-Simons formulation, the BTZ black hole solution and the AdS$_3$/CFT$_2$ correspondence. While this material is well known, it makes the review self contained and emphasizes  similarities in the construction of the Chern-Simons formulation  of ordinary gravity and higher spin gravity as well as the construction of  BTZ black hole and the higher spin black holes.  In section 3 we describe the construction of black holes carrying higher spin charge in $sl(N,R)\oplus sl(N,R)$ higher spin gravity.  The simplest case $N=3$ is discussed in detail.
 In section 4 the construction  of black holes is generalized to the \hsl $\oplus $ \hsl\ higher spin gravity. An important check of the proposal is performed by comparing the asymptotic form of the gravity partition function with  CFT. In section 5 we conclude with a list of open problems and directions for future research.

\newsec{Black holes in three dimensional gravity}

Three dimensional gravity has no propagating degrees of freedom; the Einstein equations fix the metric to be locally flat in the absence of a cosmological constant, or locally  (A)dS in the presence of a (negative) positive cosmological constant.  Nevertheless, the theory is far from trivial.  In particular, the case with negative cosmological constant admits black holes and figures importantly in the AdS/CFT correspondence.   In this section we review the main aspects of this story, and set the stage for the generalization to the higher spin theory.  We begin by working in the metric formulation, and then show how to recover the same results in the Chern-Simons formulation.  Some relevant pedagogical references include \refs{\BanadosGG,\KrausWN,\CampoleoniZQ}.

\subsec{Action, boundary conditions, and Virasoro symmetry}

The action for pure gravity with negative cosmological constant is %
\eqn\pkaa{I=  {1 \over 16\pi G } \int \! d^3 x \sqrt{g} \,(R-{2
\over \ell^2}) +I_{\rm bndy}~.}
We work in Euclidean signature.  The boundary terms contained in $I_{{\rm bndy}}$ are needed to ensure a proper variational principle, and will be displayed below.

One solution of the equations of motion is AdS$_3$ in global coordinates,
\eqn\pkab{ ds^2 = (1+r^2/\ell^2)dt^2 +{dr^2 \over 1+r^2/\ell^2}+r^2 d\phi^2~.}
As noted above, this is in fact the general solution of the field equations up to coordinate transformations and global identifications.   The physical content of the theory, at least at the classical level, thus corresponds to understanding the effect of these coordinate transformations and global identifications.

To define the theory we need to impose boundary conditions at infinity. These take the form of falloff conditions on the metric components.  We'd like to allow for the most general falloff conditions compatible with a well-defined variational principle.  It is convenient to write the line element in the form
\eqn\pkac{ ds^2 =  d\rho^2 + g_{ij}(x^k,\rho)dx^i dx^j~,\quad i,j = 1,2~.}
We then demand that $g_{ij}$ takes the Fefferman-Graham \fef\ form as $\rho\rightarrow \infty$,
\eqn\pkad{ g_{ij}(x^k,\rho)dx^i dx^j = e^{2\rho/\ell}g^{(0)}_{ij}(x^k) + g^{(2)}_{ij}(x^k) + \ldots ~.}
This defines $g^{(0)}_{ij}$ as the conformal boundary metric; in the AdS/CFT correspondence the CFT lives on a space with this metric.  By allowing for a general $g^{(0)}_{ij}$ we arrive at the notion of an asymptotically, locally, AdS$_3$ spacetime.  See, e.g., \deHaroXN\ for more details.

Our variational principle is defined by the demand that the action be stationary under any variation that respects the Einstein equations and the falloff conditions \pkad, with $g^{(0)}_{ij}$ fixed.  With $g^{(0)}_{ij}$ allowed to vary, we will have the on-shell variation
\eqn\pkae{ \delta I = \half \int  \! d^2x \sqrt{g^{(0)}}\,
T^{ij} \delta g^{(0)}_{ij}~,}
which defines the boundary stress tensor $T^{ij}$ \BalasubramanianRE.  These requirements fix the boundary terms in the action to be
\eqn\pkaf{ I_{\rm bndy} = {1 \over 8 \pi G}\int_{\p {\cal M}} \! d^2 x
\sqrt{g}\, \left( \Tr K  -{1 \over  \ell}\right) +  {\rm anomaly~term}~,}
where $K_{ij} = {1\over 2} \p_\eta g_{ij}$ is the extrinsic curvature.  The anomaly term is needed to cancel a logarithmic divergence, but it will not be needed here.   The integral in \pkaf\ is performed at a fixed $\eta$, whose value is taken to infinity at the end of the computation, with a finite result.  The boundary stress tensor works out to be
\eqn\pkag{T_{ij} = {1 \over 8\pi G \ell}
\left(g^{(2)}_{ij}-\Tr (g^{(2)} ) g^{(0)}_{ij}\right)
~,}
where the trace is defined by raising an index with $g^{(0)ij}$.

We now consider the case that the conformal boundary is a cylinder, as in \pkab. We work in complex coordinates, $w= \phi+it/\ell$, so that $g^{(0)}_{ww}= g^{(0)}_{\overline{w}\overline{w}}=0$ and $g^{(0)}_{w\overline{w}} = 1/2$.   The asymptotic symmetry group is obtained by considering coordinate transformations that preserve the form of $g^{(0)}_{ij}$.  This is the case for the following infinitesimal transformations \BrownNW:
\eqn\pkah{\eqalign{ w & \rightarrow w + \eps(w) - {\ell^2 \over 2} e^{-2\rho/\ell} \p_{\wb}^2 \epsb(\wb) \cr
 \wb & \rightarrow \wb + \epsb(\wb) - {\ell^2 \over 2} e^{-2\rho/\ell} \p_{w}^2 \eps(w) \cr
 \rho &\rightarrow \rho - {\ell\over 2} \Big( \p_w \eps(w) + \p_{\wb} \epsb(\wb)\Big)~,  }}
where $\eps(w)$ and $\epsb(\wb)$ are arbitrary functions.  These coordinate transformations leave $g^{(0)}_{ij}$ fixed but act nontrivially on $g^{(2)}_{ij}$, and so transform the stress tensor as
\eqn\pkai{ T_{ww} \rightarrow T_{ww}  +2 \p_w \eps(w) T_{ww} + \eps(w) \p_w T_{ww} -{c\over 24\pi} \p_w^3 \eps(w) }

with
\eqn\pkaj{ c= {3\ell \over 2G }~.}
The analogous result holds for $T_{\wb\wb}$.
This is the transformation law for a stress tensor in a two-dimensional conformal field theory, with $c$ being the central charge.

By following the rules of the AdS/CFT correspondence we can compute boundary correlation functions of the stress tensor.  Following precisely the same logic as in CFT (e.g., see \PolchinskiRQ), the conformal symmetry implied by \pkai\ leads to OPE relations\foot{Compared to typical CFT conventions, the stress tensor defined from gravity as here contains an extra factor of $2\pi$; in section 3 we switch to CFT conventions, see e.g. equation (3.8).}
\eqn\pkak{ T_{ww}(w) T_{ww}(0) \sim {c\over 8\pi^2 w^4} +{1\over \pi w^2} T_{ww}(0) + {1\over 2\pi w} \p_w T_{ww}(0)~.}
Alternatively, if we define the mode generators $L_n$ via
\eqn\pkal{\eqalign{ L_n & = \oint \! dw~ e^{-in w} T_{ww} }}
then the generators will obey the Virasoro algebra
\eqn\pkam{ [L_m,L_n] = (m-n)L_{m+n} +{c\over
12}(m^3-m)\delta_{m+n}~.}
At the level of classical gravity \pkam\ is a statement about Poisson brackets, while in the dual CFT it is an operator statement. The zero mode generators are related to mass and angular momentum as
\eqn\pkan{ L_0 = {M\ell -J \over 2}~,\quad \overline{L}_0 = {M\ell +J \over 2}~.}

\subsec{The BTZ black hole \BanadosWN}

The metric of the rotating Euclidean BTZ black is
\eqn\pkba{ ds^2 = {(r^2 -r_+^2)(r^2 -r_-^2) \over r^2
\ell^2}dt^2+{\ell^2 r^2 \over
(r^2-r_+^2)(r^2-r_-^2)}dr^2+r^2(d\phi + i{r_+  r_- \over \ell r^2}
dt)^2~.}
To obtain a real Euclidean signature metric, $r_-$ is taken to be purely imaginary.
The global identifications that lead from pure AdS$_3$ to BTZ may be found in \BanadosWN.  In Lorentzian signature, the BTZ solution is a black hole with inner/outer horizons at $r=r_\pm$.   Besides the $2\pi$ periodicity of the angular coordinate $\phi$, smoothness of the Euclidean signature metric at $r=r_+$ requires that we impose the further identification
\eqn\pkbb{ w\cong w+2\pi \tau~,\quad \tau = {i\ell \over r_+ + r_-}~,}
where $w$ is defined as before, $w= \phi + it/\ell$.  With these identifications, the conformal boundary of the BTZ black hole is seen to be a torus of modular parameter $\tau$.

The BTZ solution can be written in the Fefferman-Graham form \pkad\ as
\eqn\pkbc{\eqalign{  ds^2& =  d\rho^2 +8\pi G\ell \Big( \Lc dw^2 +\Lcb d\wb^2 \Big)+\Big(\ell^2 e^{2\rho/\ell}+(8\pi G)^2 \Lc \Lcb e^{-2\rho/\ell}\Big) dw d\wb ~.}}
The constants $\Lc$ and $\Lcb$ are rescaled versions of the Virasoro zero modes,
\eqn\pkbd{ \Lc = {1\over 2\pi}L_0~,\quad  \Lcb = {1\over 2\pi}\overline{L}_0~,}
which are in turn related to $r_\pm$ as
\eqn\pkbda{ L_0 = {(r_+-r_-)^2 \over 16 G \ell}~, \quad
\overline{L}_0 = {(r_++r_-)^2 \over 16 G \ell}~.}
Other useful relations are
\eqn\pkbdb{ \tau  = {ik \over 2}{1\over \sqrt{2\pi k \Lc}}~,\quad
\overline{\tau}= {-ik\over 2}{1\over \sqrt{2\pi k \Lcb} }~.}

The area law gives the black hole entropy as
\eqn\pkbe{ S= {A\over 4G} = 2\pi \sqrt{{c\over 6}L_0} + 2\pi \sqrt{{c\over 6}\overline{L}_0}~,}
which exhibits the famous fact that the entropy is given by Cardy's formula \StromingerEQ.

It is illuminating to obtain the entropy from the Euclidean action, which is related to the thermodynamic quantities as
\eqn\pkbf{ -I = S + 2\pi i \tau L_0 - 2\pi i \overline{\tau} \overline{L}_0~,}
Computing the Euclidean action directly for the BTZ black hole will recover \pkbe.  We can instead use modular invariance to simplify the computation. In three dimensional gravity, modular transformations are coordinate transformations that act on the boundary so as to change the modular parameter as
\eqn\pkbg{ \tau \rightarrow \tau'= {a\tau +b \over c\tau +d}~,\quad ad-bc=1~.}
In fact, we can find a coordinate transformation such that the BTZ metric goes over into the pure AdS$_3$ metric \pkab\ with modular parameter $\tau' = -1/\tau$ \MaldacenaBW.  The Euclidean action of this solution is easily worked out as

\eqn\pkbh{ I = {i\pi  c\over 12}(\tau'-\overline{\tau}') =-{i \pi c\over 12}\left({1\over \tau}-{1\over \taub}\right)~.}
Since the Euclidean action is invariant under coordinate transformations, this is therefore the action for the BTZ black hole. Writing the result in the form \pkbf\ yields the entropy formula \pkbe.

 A major virtue of  the modular transformation approach is that it can be applied to cases where the area law is no longer valid, such as when higher derivative terms are present in the action.  All such corrections can be absorbed into the central charge $c$, and taking this into account Cardy's formula continues to be valid \KrausVZ.

\subsec{Charged BTZ black holes}

Our main focus in this review is the construction and  study of black holes endowed with higher spin charge.  A simpler version of this problem, free from the complications of higher spin gravity, involves black holes carrying a spin-1 charge.   These have a rather simple description when we add to our theory a pure Chern-Simons U(1) gauge field.

To our previous action we now add
\eqn\pkca{ I_{CS}= {i\hat{k}\over 8\pi }\int A dA - {\hat{k}\over 16\pi} \int\! d^2x \sqrt{g} A^i A_i~,}
where $A$ is a 1-form.  The second term is a boundary term (the metric that appears in this term is that induced on the boundary), introduced for the same reasons as in the pure gravity case \KrausNB.   We impose boundary conditions such that in Fefferman-Graham coordinates $A_\rho$ vanishes and
\eqn\pkcb{ A_i(\rho,x^j) = A^{(0)}_i (x^j) + \ldots~,\quad \rho \rightarrow \infty~.}
See \AndradeSX\ for a discussion of more general boundary conditions, including the case where a Maxwell term is present.
The on-shell variation of the action then takes the form
\eqn\pkcc{ \delta I_{CS}= {i\over 2\pi} \int \! d^2 w \sqrt{g^{(0)}}J^{\wb} \delta A^{(0)}_{\wb}}
where the boundary current is
\eqn\pkcd{ J_w = {1\over 2} J^{\wb}={i\hat{k}\over 2} A^{(0)}_w~.}
$A^{(0)}_{\wb}$ plays the role of a source conjugate to the current $J_w$. In particular, to incorporate $A^{(0)}_{\wb}$ on the CFT side of the AdS/CFT correspondence we add to the CFT action the term
\eqn\pkcda{ I_{{\rm CFT}} \rightarrow I_{{\rm CFT}} + {i\over 2\pi} \int\! d^2 w J^{\wb} A^{(0)}_{\wb}~.}
Note that in the absence of a source the current is holomorphic, $\p_{\wb} J_w = {i\hat{k}\over 2} \p_{\wb} A^{(0)}_w = 0$, where in the last step we used the Chern-Simons equations of motion, namely that the field strength vanishes.   An anti-holomorphic current is obtained by starting with \pkca, but flipping the sign of the first term.  The modes of the current, defined as,
\eqn\pkcda{ J_n = \oint\! {dw \over 2\pi i} e^{-inw} J_w }
furnish a $U(1)$ Kac-Moody algebra \KrausNB.

Since the field strength vanishes by the Chern-Simons equations of motion, the gauge field is locally pure gauge.  However, in the presence of a black hole the gauge field can have a nontrivial holonomy around the horizon, and this translates into an electric charge carried by the black hole.  The BTZ boundary torus has two non-contractible cycles, defined by the two identifications $w\cong w+2\pi$ and $w\cong w+2\pi \tau$.   The former is noncontractible when extended into the bulk (it goes around the horizon), while the latter is contractible.  A smooth gauge field must have trivial holonomy around the contractible cycle, which imposes the condition
$\tau A_w + \overline{\tau} A_{\wb}=0$.   Subject to this constraint, we can choose an arbitrary flat connection, and then the black hole charge is given by $J_0$.

The black hole entropy is still given by the usual area law, and the BTZ metric is clearly unaffected by the addition of a flat U(1) connection.  However, the boundary term appearing in \pkca\ depends on the metric, and this induces a shift in the stress tensor.  So the formula for the entropy is modified when expressed in terms of the Virasoro and U(1) charges, and is given by \KrausNB\
\eqn\pkce{S=2\pi \sqrt{{c\over 6}(L_0 - {1\over \hat{k}} J_0^2)} + 2\pi \sqrt{{c\over 6}\overline{L}_0}~,}
where we include just a holomorphic current.

\subsec{Chern-Simons formulation of three dimensional AdS gravity}

An  alternative formulation of three dimensional gravity is as a Chern-Simons theory \refs{\AchucarroVZ,\WittenHC}.  We now review this story and show how to recover the results found in the metric formulation.

We consider the action
\eqn\pkda{ S = S_{CS}[A] - S_{CS}[\Ab]}
where

\eqn\pkdb{  S_{CS}[A] ={k\over 4\pi} \int\! \Tr \left(A\wedge dA +{2\over 3} A\wedge A \wedge A\right) ~.}
The 1-forms $A$ and $\Ab$ take values in the Lie algebra of $SL(2,R)$. The Chern-Simons level $k$ will eventually be related to ratio of the AdS$_3$ radius $\ell$ and the Newton constant $G$ as
\eqn\pkdc{ k = {\ell \over 4G}~.}
For convenience, we henceforth set $\ell =1$.

The Chern-Simons equations of motion correspond
to vanishing field strengths,
\eqn\pkdd{ F = dA + A\wedge A =0~,\quad \overline{F}= d\Ab+\Ab\wedge \Ab =0~. }
To relate these to the  Einstein equations we introduce a vielbein
$e$ and spin connection $\omega$ as
\eqn\pkde{ A = \omega+e ~,\quad \Ab= \omega -e~. }
The flatness conditions \pkdd\ then become Einstein's equations in first order form.   The metric is obtained from the vielbein in the usual fashion
\eqn\pkdf{ g_{\mu\nu} =2 \Tr(e_\mu e_\nu)~.}

To define asymptotically AdS boundary conditions it is convenient to choose an explicit basis for the $sl(2,R)$ generators.  We take
\eqn\pkdg{ L_1 =\left(\matrix{0&0\cr -1&0}\right)~,\quad L_{-1} =\left(\matrix{0&1\cr 0&0}\right)~,\quad L_0 =\left(\matrix{1/2&0\cr 0&-1/2}\right) }
which obey
\eqn\pkdh{\eqalign{&[L_i,L_j] = (i-j)L_{i+j}~.}}
%
The BTZ metric \pkbc\ is reproduced starting from
\eqn\pkdi{\eqalign{ A& =\big( e^\rho L_1 -{2\pi \Lc\over k}  e^{-\rho}L_{-1}\big)dz + L_0 d\rho \cr
 \Ab& =\big( e^\rho L_{-1} -{2\pi \Lcb\over k}  e^{-\rho}L_{1}\big)d\zb - L_0 d\rho}}
where we're now using $z = \phi + it$ rather than $w$ as our complex coordinate.  Noting that the dependence on the BTZ charges $(\Lc,\Lcb)$ arises purely through subleading terms at large $\rho$, we define an asymptotically AdS$_3$ connection to be one which differs from \pkdi\ by terms that go to zero for large $\rho$.

Upon using the freedom to make gauge transformations (see \CampoleoniZQ\ for a proof), asymptotically
AdS$_3$ connections can be taken to have the form
\eqn\pkdj{\eqalign{ A & = b^{-1} a(z) b + b^{-1} db~,\quad
\Ab  = b \ab(\zb) b^{-1} + b db^{-1} }}
with
\eqn\pkdk{b = e^{\rho L_0}~,}
and
\eqn\pkdl{\eqalign{ a(z)&=\Big( L_1 -{2\pi \over k} \Lc(z) L_{-1} \Big)dz \cr
\ab(\zb)&= \Big( L_{-1} -{2\pi \over k} \Lcb(\zb) L_{1}\Big)d\zb~. }}
The coefficient functions $(\Lc(z), \Lcb(\zb))$ are nothing but the components of the boundary stress tensor,
\eqn\pkdm{ T_{zz} = \Lc~,\quad T_{\zb\zb}= \Lcb~.}
This can be established in a variety of ways.  One approach is to work out the metric corresponding to \pkdl\ and then read off the stress tensor components from \pkag.   Alternatively, we can study the asymptotic symmetries directly in the Chern-Simons formulation, and use this to identify $(\Lc(z), \Lcb(\zb))$ as the currents associated with conformal symmetries \refs{\CampoleoniZQ,\GutperleKF}.

Focussing on the unbarred connection, infinitesimal gauge transformations act as
\eqn\pkdmb{ \delta A = d\Lambda + [A,\Lambda]~.}
To preserve the structure in \pkdj\ we take
\eqn\pkdma{ \Lambda = b^{-1} \lambda(z) b~.}
Further, the form of $a(z)$ is \pkdl\ is preserved by taking
\eqn\pkdn{ \lambda(z) = \eps(z) L_1 - \p_z \eps(z) L_0 + \left({1\over 2} \p_z^2 \eps(z) -{2\pi \over k} \Lc \eps(z)\right)L_{-1}~,}
which acts as
\eqn\pkdo{ \delta \Lc = \eps(z) \p_z \Lc +2 \p_z \eps(z) \Lc -{k\over 4\pi} \p_z^3 \eps(z)~.}
Comparing with \pkai\   shows that  $\Lc$ is indeed transforming as a stress tensor under the asymptotic symmetries.

Now let us return to the BTZ solution as expressed by the connections \pkdi.  In the Chern-Simons formulation the metric is a derived concept, and the physical content of a solution is far from obvious just by inspecting the connection.  For instance, the Euclidean BTZ spacetime \pkbc\ smoothly caps off at the horizon, $e^{2\rho_+} = {2\pi \over k}\sqrt{\Lc \Lcb}$, whereas from the Chern-Simons perspective there is no obvious reason we should restrict the range of $\rho$.

In the metric formulation the relation between the charges $(\Lc, \Lcb)$ and the modular parameter $\tau$ is fixed by demanding the absence of a conical singularity at the horizon.  We need a corresponding condition expressed in Chern-Simons language; i.e. in terms of the connection and not the metric.   The idea is to focus on the holonomy of the connection around the Euclidean time circle.   This holonomy should be trivial if we want to say that this circle smoothly closes off at the horizon.  The holonomy of $A$ is, up to conjugation by $b$,
\eqn\pkdp{   H= e^{2\pi \tau a_z } ~. }
 $2\pi \tau a_z$ has eigenvalues $ \pm \sqrt{8\pi^3 \Lc \over k}\tau$.  If we use the relations \pkbdb\ we find that the eigenvalues are $\pm i\pi$, so that,  $H={\rm diag}(-1,-1)$.  Since this  $H$ is in the center of $SL(2,R)$, it indeed represents a trivial holonomy.

 \subsec{Black hole entropy in the Chern-Simons formulation}

Next we turn to computing the black hole entropy.  As reviewed in the next section, with suitable attention paid to boundary terms, it is possible to compute the black hole free energy via the Euclidean Chern-Simons action  \BanadosUE, but a more direct approach is to use the first law of thermodynamics, which is how we proceed here.   We can define the black hole partition function as
\eqn\pkdq{\eqalign{  Z(\tau,\overline{\overline{\tau}}) &  = \Tr \left[e^{4\pi^2 i \tau \hat{\Lc} - 4\pi^2 i \overline{\tau} \hat{\Lcb} } \right] \cr
& =  e^{S +4\pi^2 i \tau \Lc - 4\pi^2 i \overline{\tau} \Lcb }~,}}
where the operator expression written in the first line is essentially a mnemonic for present purposes, but can be thought of as the fundamental definition of the partition function in some putative
microscopic theory, while the second line represents the saddle point approximation, which is what we use here.   Factors of $2\pi$ arise from the rescaling in \pkbd\ as well as from the $2\pi \tau$ identification.

From the definition of $Z$ it follows that
\eqn\pkdr{ \Lc = -{i\over 4\pi^2} { \p \ln Z \over \p \tau}~,\quad  \Lcb = {i\over 4\pi^2} { \p \ln Z \over \p \overline{\tau} }}
and
\eqn\pkds{ S = \ln Z - 4\pi^2 i \tau \hat{\Lc} + 4\pi^2 i \overline{\tau} \hat{\Lcb}~.}
Since expressions for $(\Lc, \Lcb)$ in terms of $(\tau, \overline{\tau})$ have already been fixed by the holonomy condition, namely as \pkbdb, we can integrate \pkdr\ to find $\ln Z$, and then use $\pkds$ to compute $S$.  Doing so, we of course reproduce the previous result \pkbe.

We now consider a more general case in which additional charges are present.    Consider a charge $Q$ with corresponding conjugate potential $\alpha$.  In such a case we can define the partition function
\eqn\pkdt{ Z(\tau,\alpha) = \Tr \left[e^{4\pi^2 i (\tau \hat{\Lc}+\alpha \hat{Q})}\right]= e^{S + 4\pi^2 i (\tau {\Lc}+\alpha {Q})}~,}
where we suppress the dependence on $\overline{\tau}$ for clarity. Now, suppose we are given a black hole solution and we wish to determine $Z$, and hence $S$.    As in the case above, a first step is to determine expressions for $\Lc$ and $Q$ in terms of $\tau$ and $\alpha$.    These relations will be fixed by demanding appropriate smoothness conditions at the horizon.    There is an important consistency condition that any such smoothness condition must respect if an underlying partition function is to exist.  Namely, since
\eqn\pkdu{ \Lc =  -{i\over 4\pi^2} { \p \ln Z \over \p \tau}~,\quad Q= -{i\over 4\pi^2} { \p \ln Z \over \p \alpha}~,}
the existence of $Z$ implies
\eqn\pkdv{  {\p \Lc \over \p \alpha} = {\p Q \over \p \tau}~.}
Conversely, given expressions  $\Lc(\tau,\alpha)$ and $Q(\tau,\alpha)$,  in order for $Z$ to exist the integrability condition \pkdv\ must be obeyed.   In the case of higher spin black holes there are some subtleties in defining what precisely is meant by a solution being smooth; for black holes an important check on any proposal is that it leads to a charge assignments compatible with the integrability condition.

We will determine the thermodynamic properties of higher spin black holes using the approach just outlined.  The alternative approach \BanadosUE\  (see also \PerezCF) based on evaluating the Euclidean  Chern-Simons action with appropriate boundary terms leads to the same final result.


\newsec{Black holes in $sl(N,R) \oplus sl(N,R)$ higher spin gravity}
In the last section we reviewed that Einstein gravity with a negative cosmological constant can be reformulated as a $sl(2,R) \oplus sl(2,R)$ Chern-Simons theory. In a series of recent papers generalizations to Einstein gravity with a negative cosmological constant coupled to higher spin fields were discussed. While there is no straightforward generalization within the metric formulation we can easily generalize the Chern-Simons formulation. In particular  in  \refs{\BlencoweGJ,\BergshoeffNS} it was shown that a $sl(N,R)\oplus sl(N,R)$ Chern-Simons theory corresponds to Einstein gravity coupled to $N-2$ symmetric tensor fields of spin $s=3,4, \ldots, N$. For simplicity, we restrict ourselves to $N=3$ corresponding to Einstein gravity with a spin-3 field and study this theory in more detail. In particular, we construct black hole solutions with spin-3 charge.

\subsec{Review of $sl(3,R) \oplus sl(3,R)$ higher spin gravity and $W_3$ symmetry}
In this section we first review the formulation of three dimensional higher spin gravity in terms of $sl(3,R) \oplus sl(3,R)$ Chern-Simons theory. As in the case of Einstein gravity with a negative cosmological constant we use the action $S = S_{CS}[A] - S_{CS}[\bar{A}]$ where the Chern-Simons action is given by equation \pkdb. The 1-forms $A$ and $\Ab$ take values in the Lie algebra $sl(3,R)$ and satisfy the flatness conditions
\eqn\maaa{ F = dA + A\wedge A =0~,\quad \overline{F}= d\Ab+\Ab\wedge \Ab =0~. }
Whenever necessary, we use the following basis of $sl(3,R)$ generators\foot{Compared to ref \CampoleoniZQ, we set $\sigma = -1.$}
\eqn\maab{ \eqalign{ L_1 & = \left(\matrix{0&0&0 \cr 1&0&0 \cr 0&1&0}\right),\quad L_0= \left(\matrix{1&0&0 \cr 0&0&0 \cr 0&0&-1}\right),\quad  L_{-1} = \left(\matrix{0&-2&0 \cr 0&0&-2 \cr 0&0&0}\right),\cr & \cr
W_2 &= \left(\matrix{0&0&0 \cr 0&0&0 \cr 2&0&0}\right),\quad W_1 = \left(\matrix{0&0&0 \cr 1&0&0 \cr 0&-1&0}\right),\quad  W_0 = {2\over 3} \left(\matrix{1&0&0 \cr 0&-2&0 \cr 0&0&1}\right),\cr & \cr
W_{-1} &= \left(\matrix{0&-2&0 \cr 0&0&2 \cr 0&0&0}\right),\quad W_{-2} = \left(\matrix{0&0&8 \cr 0&0&0 \cr 0&0&0}\right), }}
which satisfy the following commutation relations
\eqn\maac{\eqalign{ [L_i ,L_j] &= (i-j)L_{i+j} ~, \cr
[L_i, W_m] &= (2i-m)W_{i+m}~, \cr
[W_m,W_n] &= -{1 \over 3}(m-n)(2m^2+2n^2-mn-8)L_{m+n}~. }}
Note that the generators $L_i$ generate a $sl(2,R)$ subalgebra of $sl(3,R).$ Under this $sl(2,R)$ the generators $W_m$ form a spin two multiplet. This is called the principal embedding of $sl(2,R)$ into $sl(3,R)$ which we use in this section. Note that this is not the only possible inequivalent embedding of $sl(2,R)$ into $sl(3,R)$: instead of $(L_1, L_0, L_{-1})$ we can consider the $sl(2,R)$ subalgebra generated by $(W_2/4, L_0/2, - W_{-2}/4).$

Instead of using $A$ and $\Ab$ we can also determine the vielbein $e$ and the spin connection $\omega$ according to equation \pkde. Expanding $e$ and $\omega$ in a basis of 1-forms $dx^\mu$, the spacetime metric $g_{\mu\nu}$  and spin-3 field $\varphi_{\mu\nu\gamma}$
are identified as\foot{Note that we used a different convention for $L_i$ in this section. In particular the normalization of $L_0$ is different since in this section $\Tr (L_0 L_0) = 2$ while in section 2 we used $sl(2,R)$ generators with $\Tr (L_0 L_0)=1/2.$ Therefore the normalization of $g_{\mu\nu}$ is different from the one used in equation \pkdf.}
\eqn\maad{ g_{\mu\nu} = {1\over 2} \Tr (e_\mu e_{\nu} )~,\quad \varphi _{\mu\nu\gamma} ={1\over 3! } \Tr ( e_{(\mu} e_\nu e_{\gamma)})}
where $\varphi_{\mu\nu\gamma}$ is totally symmetric as indicated. Restricting to the $sl(2,R)$ subalgebra generated by $L_i$, the flatness conditions \maab\ can be seen to be equivalent to Einstein's equations for the metric $g_{\mu\nu}$ with a torsion free spin-connection \CampoleoniZQ.  More generally, we find equations describing a consistent coupling of the metric to the spin-3 field.

Acting on the metric and spin-3 field, the $sl(3,R)\oplus sl(3,R)$  gauge  symmetries of the Chern-Simons theory turn into diffeomorphisms  along with
spin-3 gauge transformations (the Chern-Simons gauge transformation also include frame rotations, which leave the metric and spin-3 field invariant).  Under diffeomorphisms, the metric and spin-3 field transform according to the usual tensor transformation rules.   The spin-3 gauge transformations are less familiar, as they in general act nontrivially on both the metric and spin-3 field. It is worth noting, though, that if we ignore the spin-3 gauge invariance, then we can view the theory as a particular diffeomorphism invariant theory of a metric and a rank-3 symmetric tensor field. In \CampoleoniHP\ an action for a  metric-like formulation of the $sl(3,R)\oplus sl(3,R)$  Chern-Simons theory was formulated which is valid up to quadratic order in the spin-3 field.

Let us consider the following connection:
\eqn\maae{\eqalign{ A(z)&= \Big(e^{\rho} L_1 - {2\pi \over k} e^{-\rho} \Lc(z) L_{-1} -{\pi \over 2k} e^{-2\rho} \Wc(z) W_{-2} \Big)dz +L_0 d\rho  \cr
\Ab(\zb)&= \Big(e^\rho L_{-1} -{2\pi \over k} e^{-\rho} \Lcb(\zb) L_{1} -{\pi \over 2k} e^{-2\rho} \Wcb(\zb) W_{2} \Big)d\zb -L_0 d\rho ~.}}
Setting $\Wc(z)=0$ and $\Wcb(\zb)=0$ we recover the connection \pkdi. Moreover, since $A_{\zb}=0, A_\rho= L_0$ and
\eqn\maaf{A- A_{AdS} \sim {\cal O}(1)~~~ {\rm as }~~ \rho \rightarrow \infty}
with analogous conditions for $\Ab,$ the connection \maae\ is asymptotically AdS \CampoleoniZQ.

As explained in section 2 it is convenient to use a gauge transformation of the form \pkdj\ and \pkdk\ to obtain
\eqn\maag{\eqalign{ a(z)&= \Big(L_1 -{2\pi \over k}\Lc(z) L_{-1} -{\pi \over 2k} \Wc(z) W_{-2} \Big)dz  \cr
\bar{a}(\zb)&= \Big(L_{-1} -{2\pi \over k}\Lcb(\zb) L_{1} -{\pi \over 2k} \Wcb(\zb) W_{2} \Big)d\zb ~. }}
The coefficients $\Lc(z)$ and $\Lcb(\zb)$ are the components of the energy momentum tensor as identified in equation \pkdm. The new coefficient functions $\Wc$ and $\Wcb$ correspond to the spin-3 currents as we will see now by analyzing the asymptotic symmetry algebra.

Given the connection \maae, the asymptotic symmetry algebra is obtained by finding the most general gauge transformation \pkdm\ preserving the asymptotic conditions \maae. The functions appearing in the connection \maae,  $\Lc,\Lcb, \Wc$ and $\Wcb$, transform under these gauge transformations.  Expanding in modes, one thereby arrives at two copies of the classical $W_3$ algebra \ZamolodchikovWN.

Alternatively (see section 4 of \GutperleKF\ for details), one can translate these variations into an operator product expansion for the symmetry currents,\foot{Beware the $2\pi$ difference in stress tensor normalization relative to the convenient gravity convention, as pointed out in footnote 2.}
\eqn\maah{\eqalign{ T(z) T(0)& \sim  {3k \over z^4} + {2\over z^2}  T(0) + {1\over z} \p T(0)  \cr
T(z) \Wc(0) & \sim  {3\over z^2} \Wc(0) +{1\over z}\p \Wc(0) \cr
\Wc(z) \Wc(0) & \sim
 -{5  k \over \pi^2} {1\over z^6}-{5 \over \pi^2} {1\over z^4} T(0)  -{5 \over 2\pi^2}{1\over z^3}\p T(0) -{3 \over 4\pi^2}{1\over z^2} \p^2 T(0) -{1 \over 6\pi^2} {1\over z} \p^3 T(0) \cr
&\quad -{8  \over 3k\pi^2}{1\over z} T(0)  \p T(0) -{8  \over 3k\pi^2}{1\over z^2} T(0)^2  }}
and the result is the classical version of the $W_3$ algebra. In order to write the $W_3$ algebra as in \ZamolodchikovWN\ we identify the central charge $c$ as $c=6k$ and we define the operator $V$ used in \ZamolodchikovWN\ by $V = {{2\pi i}\over{\sqrt{10}}} {\cal W}$. The $T\Wc$ OPE identifies $\Wc$ as a spin-3 current, i.e., as a dimension $(3,0)$ primary operator. The same analysis for the barred connection gives rise to an anti-holomorphic $W_3$ algebra with a dimension $(0,3)$ current.

There is another way to show that $sl(3,R) \oplus sl(3,R)$ Chern-Simons theory give rise to ${\cal W}_3$ symmetry algebra. Instead of considering the asymptotic symmetry algebra we consider the Ward identities which are the bulk field equations for spin-3 gravity in the presence of sources for the spin-3 operators.

For simplicity let us consider the holomorphic part only corresponding to the connection $a.$ We consider a chemical potential $\mu(z,\zb)$ for the spin-3 operator which is given by a term $-\mu(z,\zb) W_2 d\zb$ in the connection $a.$ We will justify this assignment {\it a posteriori} by deriving the Ward-Identities. For a consistent ansatz we do not only need a term of the form $-\mu W_2 d\zb$ in the connection but we have to consider
\eqn\maai{\eqalign{ a&= \Big(L_1 -{2\pi \over k}\Lc L_{-1}- {\pi \over 2k} \Wc W_{-2}  \Big) dz \cr
& \quad -\Big( \mu W_2 + w_1 W_1 + w_0 W_0 + w_{-1}W_{-1} + w_{-2}W_{-2}+  w_{-3} L_{-1} \Big)d\zb ~. }}
The flatness condition determines $w_i$ as a function of $\mu, \Lc, \Wc$ and $z$-derivatives thereof. Moreover $\Lc$ and $\Wc$ are subject to
\eqn\maaj{ \eqalign{ \p_{\zb} \Lc &= 3\Wc  \p_z \mu + 2   \p_z \Wc \mu \cr
\p_{\zb} \Wc & = {k \over 12 \pi} \p_z^5 \mu -{2 \over 3}  \p_z^3 \Lc \mu - 3 \p_z^2 \Lc \p_z \mu -5 \p_z \Lc \p_z^2 \mu -{10\over 3}  \Lc  \p_z^3 \mu +{64\pi \over 3k} \Big( \Lc  \mu  \p_z \Lc +\Lc^2 \p_z \mu \Big) ~.}}
These equations are the Ward identities as we will see. Note that $\Lc$ and $\Wc$ are now $\zb$ dependent which can be traced back to the singular terms in the OPE \maah. Let us understand this in more detail. Therefore we compute
\eqn\maak{ \p_{\zb} \langle T(z,\zb) \rangle_\mu \, , \qquad \p_{\zb} \langle \Wc(z,\zb) \rangle_\mu  }
where $\langle \cdots \rangle_\mu$ denotes an insertion of $e^{\int \mu \Wc}$ inside the expectation value.   Expanding in powers of $\mu$, and using
\eqn\maal{ \p_{\zb} \left({1\over z}\right)  = 2\pi \delta^{(2)}(z,\zb)}
as well as the OPE's \maah\ we obtain \maaj\ if we identify $T = - 2\pi \Lc.$ This justifies {\it a posteriori} the AdS/CFT dictionary for introducing the chemical potential. Moreover, we have seen that the Ward identities can be derived from the equations of motion of spin-3 gravity.

So far we considered only the principal embedding of $sl(2,R)$ into $sl(3,R)$ giving rise to $W_3$ as asymptotic symmetry. As already explained there is another inequivalent embedding of $sl(2,R)$ into $sl(3,R)$. This embedding gives rise to another $W$-algebra known as $W_3^{(2)}$, sometimes  referred to as the Polyakov-Bershadsky algebra \refs{\PolyakovDM,\BershadskyBG}. More details can be found in section 2.3 of \AmmonNK\  and \CastroBC.

\subsec{Black holes with higher spin charge}
Let us now consider black holes with higher spin charge within $sl(3,R) \oplus sl(3,R)$ Chern-Simons theory. In \GutperleKF\ the following solution was proposed to represent black holes carrying spin-3 charge:
\eqn\maba{\eqalign{ a&= \Big(L_1 -{2\pi \over k}\Lc L_{-1} -{\pi \over 2k} \Wc W_{-2} \Big)dz  \cr
& \quad -\mu \Big(W_2 -{4\pi \Lc \over k}W_0 +{4\pi^2 \Lc^2 \over k^2} W_{-2} +{4\pi \Wc \over k} L_{-1}\Big)d\zb    \cr
 \bar{a}&= -\Big(L_{-1} -{2\pi \over k}\Lcb L_{1} -{\pi \over 2k} \Wcb W_{2} \Big)d\zb \cr
 & \quad -\mub \Big( W_{-2} -{4\pi \Lcb \over k}W_0 +{4\pi^2 \Lcb^2 \over k^2} W_{2} +{4\pi \Wcb \over k} L_{1}\Big) dz ~. }}
The structure of this solution is easy to understand. Let us focus on the $a$-connection. As in \maag, to add energy and charge density to the
${\cal W}_3$ vacuum we should add to $a_z$ terms involving $L_{-1}$ and $W_{-2}$, as seen in the top line of \maba.   For black holes, which represent states of thermodynamic
equilibrium, the energy and charge should be accompanied by their
conjugate thermodynamic potentials, which are temperature and spin-3 chemical potential.  We incorporate the former via the periodicity of
imaginary time, while the latter correspond to a $\mu W_2$ term in $a_{\zb}$.  The remaining $a_{\zb}$ terms appearing in \maba\ are then fixed by the equations of motion. In fact we can rewrite the component $a_{\zb}$ of the connection \maba\ as
\eqn\mabb{a_{\zb} = - 2 \mu \Big[ (a_z)^2 - {1\over 3} \Tr (a_z^2) \Big] \, .}
In order to solve the flatness condition \maaa\ we have to show $[a_z, a_{\zb}]=0$ which is indeed satisfied by \mabb. The ansatz for the connection $ a_{\zb} $ given in \mabb\ can be generalized to more complicated cases, including black holes in \hsl\  higher spin theories.

Let us from now on restrict ourselves to the nonrotating case
\eqn\mabc{ \Lcb = \Lc~,\quad \Wcb=-\Wc~,\quad \mub = -\mu ~,}
and study the connection \maba\ in more detail. If we set $\mu = \Wc=0$, then the connections become those corresponding to a BTZ black hole asymptotic to the ${\cal W}_3$ vacuum. From the standpoint of this CFT, the $\mu W_2 d\zb$ term represents a chemical potential for spin-3 charge, and the $\Wc W_{-2} dz$ term gives the value of the spin-3 charge. This solution is therefore interpreted as a generalization of the BTZ black hole to include nonzero spin-3 charge and chemical potential.

The general non-rotating solution, i.e. the connection \maba\ imposing the conditions \mabc, can be thought of as depending on four free parameters:  three of these are  $(\Lc, \Wc, \mu)$, and the fourth is the inverse  temperature $\beta$, corresponding to the periodicity of imaginary time, $t \cong t+i\beta$.  However, we expect that there should only be a two-parameter family of physically admissible solutions: once one has specified the temperature and chemical potential the values of the energy and charge should be determined thermodynamically as a function of $\alpha$ and $\tau$ where $\tau$ is the modular parameter of the torus given by
\eqn\mabd{\tau = {{i \beta}\over{2\pi}}~,}
and $\alpha$ is the potential related to the chemical potential $\mu$ by
\eqn\mabe{\alpha = \bar{\tau} \mu\, , \qquad \bar{\alpha} = \tau \bar{\mu}~.}
Which conditions do the functions $\Lc = \Lc(\tau,\alpha)$ and $\Wc = \Wc(\tau,\alpha)$ satisfy? As discussed in section 2.4 for the uncharged BTZ solution the relation between the energy and the temperature is obtained by demanding the absence of a conical singularity at the horizon in Euclidean signature. The analogous procedure in the presence of spin-3 charge is more subtle. For the charged spin-3 black hole solution, we would like to impose the following three conditions:\medskip
{\it (i)} The Euclidean geometry is smooth and the spin-3 field is nonsingular at the horizon.\medskip
{\it (ii)} In the limit $\mu\to 0$ the solution goes smoothly over to the BTZ black hole. In particular, we want that $\Wc\rightarrow 0.$\medskip
{\it (iii)} The charge assignment $\Lc = \Lc(\tau,\alpha)$ and $\Wc = \Wc(\tau,\alpha)$ should arise from an underlying partition function and therefore, as discussed in section 2.5, the charge assignment should obey the integrability condition
\eqn\mabf{ {\p \Lc \over \p \alpha} = {\p \Wc \over \p \tau}~.}
Let us first discuss the condition {\it (i)}. The metric \maad\ associated with the gauge connection \mabb\ is given by
\eqn\mabd{ ds^2 = d\rho^2  - {\cal F}(\rho) dt^2 +{\cal G}(\rho) d\phi^2}
with
\eqn\mabe{\eqalign{
{\cal F}(\rho) &= \big( 2\mu e^{2\rho}+{\pi \over k}\Wc e^{-2\rho}-{8\pi^2 \over k^2}\mu \Lc^2 e^{-2\rho}\big)^2  +\big(e^\rho-{2\pi \over k}\Lc e^{-\rho}+{4\pi \over k}\mu \Wc e^{-\rho}\big)^2 \cr
{\cal G}(\rho) &= \big(e^\rho+{2\pi \over k}\Lc e^{-\rho}+{4\pi \over k}\mu \Wc e^{-\rho}\big)^2  \cr&+4\big( \mu e^{2\rho}+{\pi \over 2k}\Wc e^{-2\rho}+{4\pi^2 \over k^2}\mu \Lc^2 e^{-2\rho}\big)^2 +{4\over 3} \left(4\pi \over k\right)^2 \mu^2 \Lc^2~.}}
The metric reduces to BTZ in the uncharged limit, but the surprise is that for generic charge, ${\cal F}(\rho)$ and ${\cal G}(\rho)$ are both positive definite quantities. The horizon occurs where $g_{tt}$ vanishes, and since the radial dependence of the metric is simply $d\rho^2$, we need $g_{tt}$ to have a double zero at the horizon in order for the Euclidean time circle to smoothly pinch off. One solution is given by $\Wc=0$ and $\mu=0$, which is  the nonrotating  BTZ black hole. There is a second solution provided
\eqn\mabf{k+ 32 \mu^2 \pi (\mu \Wc- \Lc) =0}
is satisfied. The temperature is then determined by demanding the absence of a conical singularity at the horizon, which is located at $e^{\rho_+}= \sqrt{(2\pi\Lc -4\pi \mu \Wc )/k}$,
\eqn\mabg{\beta = 2\pi \left.\sqrt{2 \over -g_{tt}''}~\right|_{\rho=\rho_+}= {2\sqrt{2} \pi k \mu\over \sqrt{(3 k - 32\pi  \Lc \mu^2 ) ( 16 \pi \Lc \mu^2 -k)}}~.}
The two equations \mabf\ and \mabg\ determine the charge assignments $\Lc  = \Lc(\tau,\alpha)$ and $\Wc = \Wc(\tau,\alpha).$ However, these charge assignments do not satisfy the conditions {\it (ii)} and {\it (iii)}. Although these charge assignments are the only one for which the geometry of the connection \maba\ has a horizon, the charge assignments are not consistent.

For any other charge assignment, ${\cal F}$ never vanishes and there is no event horizon since this geometry possesses a globally defined timelike Killing vector.  At large positive and negative $\rho$, both ${\cal F}$ and ${\cal G}$ have leading behavior $e^{4|\rho|}$, corresponding to an AdS$_3$ metric of radius $1/2$, and so the metric \mabe\ describes a traversible wormhole connecting two asymptotic AdS$_3$ geometries.

One may therefore ask what, if anything, this solution has to do with black holes. It is at this point that we should remember that the metric of the spin-3 theory is not invariant under higher spin gauge transformations and the connection \maba\ after a suitable gauge transformation represents a smooth black hole whose charge assignment satisfies the conditions  {\it (ii)} and {\it (iii)}. In order to find this charge assignment, we have to consider the holonomy $\omega$ around the time circle which is given by
\eqn\mabh{ \omega = 2\pi (\tau a_z + \bar{\tau} a_{\zb})~.}
In \GutperleKF\ it was proposed that the eigenvalues of the holonomy $\omega$ take the fixed values $(0,2\pi i, -2\pi i)$. It is convenient to recast the conditions on its eigenvalues in the form
\eqn\mabi{\Tr(\omega^2)=-8\pi^2 ~,\quad \Tr(\o^3)=0 \, .}

We will see in section 3.3 that the holonomy condition \mabi\ gives rise to a charge assignment which satisfies the conditions  {\it (ii)} and {\it (iii)}. Moreover, in section 3.4 we find an explicit gauge transformation of the connection \maba\ such that the geometry associated with the gauge transformed connection is indeed a smooth black hole, whose smoothness enforces the holonomy condition \mabi.

\subsec{Holonomy and integrability}
In this section we evaluate the holonomy condition for the connection \maba\ and show that the consistency condtions {\it (ii)} an {\it (iii)} are indeed satisfied.

The holonomy condition \mabi\ for the connection \maba, become explicitly
\eqn\maca{\eqalign{ 0&= -2048 \pi^2 \mu^3  \Lc^3+576 \pi  k  \mu \Lc^2 -864 \pi k \mu^2  \Wc \Lc +864 \pi  k \mu^3  \Wc^2-27 k^2 \Wc \cr
0 & =256 \pi^2  \mu^2 \Lc^2  +24 \pi k  \Lc-72 \pi k \mu   \Wc +{3k^2\over \tau^2}  }}
together with the same formulas with unbarred quantities replaced by their barred versions. We have replaced $\mu$ by $\alpha$ using \mabe.

To verify the integrability conditions we proceed as follows. First solve the second equation for $\Wc$ and differentiate with respect to $\tau$ to get an expression for ${\p \Wc \over \p\tau}$ in terms of  ${\partial \Lc \over \partial\tau}$. Next, insert the solution for $\Wc$ into the first equation, and then differentiate with respect to $\alpha$, and solve to get an expression for ${\partial \Lc \over \partial \alpha}$.   Similarly, differentiate with respect to $\tau$ to get an expression for  ${\partial \Lc \over \partial \tau}$.  Substituting the latter into our previous expression for ${\p \Wc \over \p\tau}$, we find that it precisely equals ${\partial \Lc \over \partial \alpha}$, which is the desired integrability condition.

It will be convenient to define dimensionless versions of the charge and the chemical potential by
\eqn\macb{\zeta = \sqrt{k \over 32 \pi \Lc^3}\Wc~,\quad \gamma = \sqrt{2\pi \Lc  \over k}\mu ~. }
Rewriting \maca\ in terms of the quantities \macb\ we obtain
\eqn\macc{\eqalign{1728\gamma^3\zeta^2-(432\gamma^2+27)\zeta-128\gamma^3
+72\gamma&=0 \cr
\left( 1+ {16 \over 3} \gamma^2   -12 \gamma \zeta\right) \Lc -{\pi k \over 2 \beta^2}&=0~.} }
Let us solve \macc\ for the charge $\zeta$ and the inverse temperature $\beta.$

Solution of equations \maca\ for the charge $\zeta$ and inverse temperature $\beta$ yields
\eqn\macd{\eqalign{ \zeta & = {1+16\gamma^2 -\left(1-{16\over 3}\gamma^2\right)\sqrt{1+{128\over 3}\gamma^2} \over 128 \gamma^3}  \cr
\beta &= {   \sqrt{\pi k\over 2\Lc } \over \sqrt{ 1+ {16 \over 3} \gamma^2   -12 \gamma \zeta}} ~.   } }
In equation \macd\ we singled out a particular branch of $\zeta$ to ensure that for $\gamma \rightarrow 0$ also $\zeta \rightarrow 0,$ i.e. that condition {\it (ii)} defined in section 3.2 is satisfied. While the uncharged BTZ limit corresponds to $\gamma,\zeta \rightarrow 0$ the solution \macd\ limits the values $\zeta$ and $\gamma$ to
\eqn\mace{ \zeta \leq \zeta_{max} = \sqrt{{4\over 27}} \, , \qquad \gamma \leq \gamma_{max}=\sqrt{{3 \over 16}}~.}
For given $\Lc$ we thus have a maximal spin-3 charge $\Wc$ given by
\eqn\macf{ \Wc_{max}^2 = {128\pi \over 27 k }\Lc^3 ~.}
Note that these maximal values can be seen only by the holonomy conditions \maca\  and not directly from the gauge connection \maba\ or its resulting geometry.

\subsec{Finding the black hole gauge}
In this section we want to find a suitable gauge transformation to turn the connection \maba\ into a smooth black hole. In particular we will see that the smoothness conditions are indeed equivalent to the holonomy conditions \maca. The $\rho$-dependent connections corresponding to \maba\ are denoted by $A$ and $\bar{A}.$ We refer to this connection as {\it in the wormhole gauge}. We then consider new connections ${\cal A}$ and $\bar{{\cal A}}$ related to $A$ and $\bar{A}$ by $sl(3,R)$ gauge transformations:
\eqn\mada{\eqalign{ {\cal A} &= g^{-1}(\rho) A(\rho) g(\rho)+ g^{-1}(\rho) dg(\rho) \cr
\bar{{\cal A}}  &= g(\rho) \bar{A}(\rho) g^{-1}(\rho)- dg (\rho) g^{-1}(\rho)  }}
with $g(\rho) \in ~ sl(3,R)$.  The relative gauge transformation for
$\bar{{\cal A}}$ versus ${\cal A}$ is taken to maintain a non-rotating ansatz.  The
metric and spin-3 field corresponding to $(A,\Ab)$ will take the form
\eqn\madb{\eqalign{ ds^2  &= g_{\rho\rho}(\rho) d\rho^2  +g_{tt}(\rho) dt^2 + g_{\phi\phi}(\rho)d\phi^2 \cr
\varphi_{\alpha\beta\gamma} dx^\alpha dx^\beta dx^\gamma & = \varphi_{\phi \rho\rho}(\rho) d\phi d\rho^2 + \varphi_{\phi tt}(\rho) d\phi dt^2 + \varphi_{\phi\phi\phi}(\rho) d\phi^3 ~.}}
We demand that this solution describe a smooth black hole with event horizon at $\rho=\rho_+$, or at $r=0$ with
\eqn\madc{r = \rho -\rho_+ ~.}
Assuming that $g_{rr}(0) >0$, as will be the case,
this first of all  requires $g_{tt}(0) = g_{tt}'(0)=0$ and $g_{\phi\phi}(0) >0$, so that after rotating to imaginary time the metric expanded around $r=0$ will look locally like $R^2 \times S^1$:
\eqn\madd{ ds^2 \approx g_{rr}(0) dr^2 - {1\over 2}g_{tt}''(0) r^2 dt_E^2 +g_{\phi\phi}(0)d\phi^2 ~.}
In order for the metric to avoid a conical singularity at $r=0$  we need to identify $t_E \cong t_E + \beta$ with
\eqn\made{ \beta = 2\pi \sqrt{ 2g_{rr}(0) \over -g_{tt}''(0)}~.}
Having done so, we can switch to Cartesian coordinates near $r=0$ and the metric will be smooth.

The same smoothness considerations apply to the spin-3 field.
Noticing the parallel structure, we see that we should demand
$\varphi_{\phi tt}(0)  = \varphi_{\phi tt}'(0)=0$, and
\eqn\madf{ \beta = 2\pi \sqrt{ 2\varphi_{\phi rr}(0) \over -\varphi_{\phi tt}''(0)}}
with the same $\beta$ as in \made.

There is still one more condition to impose to ensure that the solution is completely smooth at the horizon.   If we work in Cartesian coordinates $(x,y)$ around $r=0$, we should demand that all functions
are infinitely differentiable with respect to both $x$ and $y$.  If this is not the case then some curvature invariant (or spin-3 quantity) involving covariant derivatives will diverge.   Given the rotational symmetry, this condition implies that the series expansion of all functions should only involve non-negative even powers of $r$.   We impose this by demanding that all functions be smooth at the horizon, and even under reflection about the horizon:
\eqn\madg{\eqalign{ g_{rr}(-r)=g_{rr}(r)~,\quad  g_{tt}(-r)&=g_{tt}(r)~,\quad  g_{\phi\phi}(-r)=g_{\phi\phi}(r) \cr
\varphi_{\phi rr}(-r)=\varphi_{\phi rr}(r)~,\quad
\varphi_{\phi tt}(-r)&=\varphi_{\phi tt}(r)~,\quad
\varphi_{\phi \phi\phi}(-r)=\varphi_{\phi \phi\phi}(r) ~. }}

In \AmmonNK\ it was shown that the symmetry conditions \madg\ can be enforced by demanding
\eqn\madh{\eqalign{ e_t(-r) &= -h(r)^{-1}e_t(r)h(r)\cr
e_{\phi}(-r) &= h(r)^{-1}e_{\phi}(r)h(r)\cr
e_{r}(-r) &= h(r)^{-1}e_{r}(r)h(r)\cr}}
with $h(r) \in~ sl(3,R)$, and similar conditions on the spin-connection. The BTZ solution has $h(r)={\bf1}$, so we can think of these conditions as a ``twist'' of the BTZ vielbein  reflection symmetries. In addition, $h(0)={\bf1}$, implying that $e_t(0)=0$, a feature of the BTZ solution that persists in the spin-3 case.\foot{Moreover, it is both surprising and convenient that the location of the horizon $r=0$ turns out to be at $\rho=\rho_+$, with $\rho_+$ given by the {\it same} expression as for BTZ: $e^{2\rho_+ } = {2\pi \Lc \over k}$.}

To gain some insight into the form of $g(r)$ and $h(r)$ we can start with the BTZ solution and then carry out the gauge transformation perturbatively in the charge.  These considerations lead us to the ansatz
\eqn\madi{\eqalign{  g(r) &= e^{F(r)(W_1-W_{-1})+G(r)L_0} \cr           h(r)&=  e^{H(r)(W_1+W_{-1})} }}
for some functions $F, G$ and $H$. Perturbation theory suggests that this ansatz gives the unique solution to our problem, although we have not proven this.  On the other hand, having assumed the ansatz \madi\ the remaining analysis definitely has a unique solution.

Even after assuming this ansatz, finding a solution that satisfies all the smoothness conditions involves a surprisingly large amount of complicated algebra requiring extensive use of Mathematica and Maple.  More details can be found in the appendix of \AmmonNK.  As we have already mentioned, a crucial point is that the solution to our problem requires that the holonomy conditions \mabi\ or equivalently \macc\ for the connection \maba\ are obeyed; therefore we can also derive the holonomy conditions by requiring the existence of a smooth black hole solution.

Here we just present the final form for the transformed metric. It will be convenient to use the variables \macb\ instead of the charges and the chemical potential and to introduce the parameter $C$ by
\eqn\madj{ \zeta = {C -1 \over C^{3/2} } ~.}
The metric takes the form \madb\ with
\eqn\madk{\eqalign{ g_{rr} &= {(C -2)(C-3) \over \left(C -2-\cosh^2(r)\right)^2} \cr
 g_{tt} &=-\left({8\pi \Lc\over k}\right)\left(  {C -3 \over C^2} \right)  {\big(a_t+b_t \cosh^2(r)\big)\sinh^2(r) \over  \left(C -2-\cosh^2(r)\right)^2} \cr
 g_{\phi\phi} &= \left({8\pi \Lc\over k}\right)\left(  {C -3 \over C^2} \right)  {\big(a_\phi+b_\phi \cosh^2(r)\big)\sinh^2(r) \over  \left(C -2-\cosh^2(r)\right)^2}+\left({8\pi \Lc\over k}\right)(1+{16\over 3} \gamma^2+12 \gamma\zeta )  ~.}}
The coefficients $a_{t,\phi}$ and $b_{t,\phi}$ are functions of $\gamma$ and $C$, given by
\eqn\madl{\eqalign{a_t &= (C-1)^2\left(4\gamma-\sqrt{C}\right)^2 ~,\cr
a_{\phi} &= (C-1)^2\left(4\gamma+\sqrt{C}\right)^2 ~,\cr
b_t &= 16\gamma^2(C-2) (C^2-2C+2)-8\gamma\sqrt{C} (2C^2-6C+5)+C(3C-4) ~,\cr
b_{\phi} &= 16\gamma^2(C-2) (C^2-2C+2)+8\gamma\sqrt{C} (2C^2-6C+5)+C(3C-4)~.\cr}}
The $t$ and $\phi$ coefficients are related by flipping the sign of $\gamma$, though this is not a bonafide sign flip of the charge, under which $C$ would also transform.

With these results in hand, we can verify that the smoothness condition of the black hole implies the holonomy condition. Demanding a smooth horizon via \made\ and \madf\ and using the definition \madj, the resulting equations are indeed equivalent to \macc\ which are the holonomy condition \mabi\ evaluated for the gauge connection \maba.

Therefore we showed that the black hole with spin-3 charge given by the connection \maba\ satisfies the three conditions {\it (i), (ii)} and {\it (iii)} defined in section 3.2. Let us now explore the black hole thermodynamics.

\subsec{Black hole thermodynamics}

Having satisfied the integrability condition, we know that if we now
compute the entropy from the partition function it is guaranteed to be consistent with
the first law of thermodynamics.   For black holes in Einstein-Hilbert
gravity, we can of course directly compute the entropy in terms of the
area of the event horizon.  But in the present context we do not know
{\it a priori} whether the entropy is related to the area in this way, in
particular due to the nontrivial spin-3 field.   We instead base our entropy
computation on demanding adherence to the first law.

Expressed as a function of $(\Lc,\Wc)$ the entropy $S$  obeys the following thermodynamic relations:
\eqn\maea{ \tau= {i\over 4\pi^2}{\p S \over \p \Lc}~,\quad   \alpha={i\over 4\pi^2}{\p S \over \p\Wc} ~.}
The analogous barred relations hold as well. Since the entropy breaks up into a sum of an unbarred piece plus a barred piece with identical structure, in the following we just focus on the unbarred part and add the two parts at the end.

We can  use dimensional analysis to write the entropy in terms of an unknown function of the dimensionless ratio $\zeta^2 \sim \Wc^2 /\Lc^3$.
With some foresight, it proves convenient to write
\eqn\maeb{ S = 2\pi \sqrt{2\pi k \Lc} f(y) }
with
\eqn\maec{ y = {27\over 2} \zeta^2 = {27 k\Wc^2 \over 64 \pi \Lc^3} ~.}
Demanding agreement with the BTZ entropy imposes $f(0)=1$.  Using \maea\ and plugging  into the second line of \maca\  we arrive at the following differential equation
\eqn\maed{ 36y \big(2 - y\big)(f')^2+f^2-1=0 ~.}
This equation also implies  the first equation in \maca. The solution with the correct boundary condition is
\eqn\maee{ f(y) = \cos \theta~,\quad \theta = {1\over 6} \arctan \left({\sqrt{y(2-y)}\over 1-y}\right)~.}
The physical range of $y$ is given by $0 \leq y \leq 2$, and we choose a branch of the arctangent such that $0 \leq \theta \leq {\pi \over 6} $.

Our final result for the entropy, including both sectors, is thus
\eqn\maef{ S = 2\pi \sqrt{2\pi k \Lc} f\Big( {27 k\Wc^2 \over 64 \pi \Lc^3}\Big)+ 2\pi \sqrt{2\pi k \Lcb} f\Big( {27 k\Wcb^2 \over 64 \pi \Lcb^3}\Big) ~.}
The function $f(y)$ takes a simple form upon plugging in for $\zeta$ as a function of $C$, using \madj. This step yields
\eqn\maeg{\theta = {1\over 6}\arctan\left(\Lambda(C)\sqrt{1-{3\over 4C}}\right)}
with
\eqn\maeh{\Lambda(C) \equiv {6\sqrt{3C}(C-1)(C-3)\over (2C-3)(C^2-12C+9)}~.}
Surprisingly, taking the cosine of this angle yields the simple expression
\eqn\maei{f(y)=\sqrt{1-{3\over 4C}}~.}
As with other quantities in our analysis, we see that the entropy is most simply expressed in terms of $C$. The extremal and zero charge limits are recovered upon inspection.

It is of course natural to wonder if the black hole entropy can be expressed in terms of a geometrical property of the horizon.  There is of course no reason to expect that the Bekenstein-Hawking area law holds, since the spin-3 field is nonzero at the horizon, and indeed one easily checks that $S \neq A/4G$.  In \CampoleoniHP\  a metric like formulation of the $sl(3,R) \oplus sl(3,R)$  higher spin theory was developed and the Wald entropy formula \WaldNT\ was applied to calculate the entropy of the black hole. At present there is a numerical discrepancy between the entropies  obtained in \maef\ and \CampoleoniHP, which remains to be understood. 

Let us also comment on the case of maximal value for the spin-3 charge \macf (for constant energy). Approaching this maximal value, $f(y)$ behaves near $y=2$ as
\eqn\maej{ f(y) \sim \sqrt{3\over 4} +{\sqrt{2}\over 12}\sqrt{2-y}+\cdots }
and $\tau$ diverges according to \maea, corresponding to vanishing chiral temperature.  On the other hand, the entropy is finite, attaining a relative value of $\sqrt{3\over 4}$ compared to the entropy at $\Wc=0$.  This behavior is to be contrasted with that of the BTZ black hole, or its charged generalizations with respect to bulk $U(1)$ gauge fields \KrausWN.  In those cases, whenever the temperature of one chiral sector goes to zero, so too does the entropy associated with that sector.
For extremal BTZ black holes the entropy is carried entirely by the sector at nonzero temperature, whereas here a zero temperature sector can contribute to the entropy.

Given the entropy, $\tau$ and $\alpha$ can be computed using \maea, and from there we compute the partition function according to
\eqn\maek{\ln Z =S+ 4 \pi^2 i \left( \tau \Lc +\alpha\Wc -\bar{\tau} \Lcb -\bar{\alpha} \Wcb \right) ~.}
This partition function should match the asymptotic behavior of the partition function of any candidate CFT dual to the higher spin theory in the bulk.

We briefly note that this entire story has been generalized without complication to the $sl(4,R) \oplus sl(4,R)$ case \TanTJ.

\subsec{Other black holes in spin-3 gravity}
In the last sections we built a higher spin black hole in spin-3 gravity. In particular we proposed a holonomy condition which allowed us to define the charge assignments in agreement with the integrability condition. We also saw that the holonomy condition is equivalent to the smoothness condition of the Euclidean black hole in the appropriate gauge. 

In \CastroFM\ another embedding of the $sl(2,R)$ subalgebra was considered, the diagonal embedding. The asymptotic symmetry algebra of the diagonal embedding is the classical ${\cal W}_3^{(2)} \times {\cal W}_3^{(2)}$ algebra. In the ${\cal W}_3^{(2)}$ algebra apart from the energy momentum tensor $T,$ we have two weight $3/2$ primaries $G_\pm$ as well as a weight one current $U.$ It turns out that in this theory it is particularly simple to construct black holes. A consistent truncation of that theory is general relativity coupled to a pair of Chern-Simons gauge fields. The connection of the BTZ black hole carrying charge under these gauge fields are given by \CastroFM\
\eqn\mafa{\eqalign{
a &= \Big[ W_2 + w W_{-2} - q W_0 \Big] dz - {\eta \over 2} W_0 d\zb \, , \cr
\bar{a} &=- \Big[ W_{-2} + w W_{2} - q W_0 \Big] d\zb + {\eta \over 2} W_0 dz \, .}}
The corresponding metric is given by
\eqn\mafb{ ds^2 = d\rho^2 - 4 \big( e^{2\rho} - w e^{-2\rho} \big)^2 dt^2 + 4 \big( e^{2\rho} + w e^{-2\rho} \big)^2 d\phi^2}
and the pair of gauge fields, $\chi$ and $\bar{\chi}$ read
\eqn\mafc{\chi = -q dz - {\eta \over 2} d\zb \, , \qquad \bar{\chi} = q d\zb +{\eta \over 2} dz \, .}
Calculating the holonomies around the Euclidean thermal circle and imposing the holonomy condition we can relate $q$ to $\eta$ as well as $w$ to the inverse temperature $\beta,$
\eqn\mafd{ \eta = 2 q \, , \qquad \beta = {1\over{8\sqrt{w}}} \, .}
Note that with this identification the Euclidean geometry is smooth and the time component of the gauge fields vanish at the horizon. Using the integrability condition we can derive a formula for the entropy. The dependence of the entropy on the charges $q$ and $w$  is exactly what it is expected from a R-charged BTZ black hole \pkce.

\subsec{Black hole entropy from the Euclidean action}

In this section we briefly discuss an approach to computing the black entropy via the Euclidean Chern-Simons action \BanadosUE.   An action based approach is very useful if one wants to include quantum corrections. In ordinary Einstein-Hilbert gravity, it is well known that after including suitable boundary terms the Euclidean action yields the correct black hole free energy \GibbonsUE.  The situation is more subtle in the Chern-Simons formulation, mainly due to the fact that the action is not invariant under gauge transformations that extend to the boundary.

The approach of \BanadosUE\ is to consider $sl(N,R)$ Chern-Simons theory on the Euclidean solid torus (recall that this is the topology of the Euclidean BTZ solution).   They focus on non-rotating solutions, with $\phi$ and $t$ being the coordinates that parametrize the non-contractible and contractible cycles of the solid torus.  Consider a connection $a = a_t dt +a_\phi d\phi$ obeying the holonomy conditions.  The authors define charges in terms of the $sl(N,R)$ Casimir operators as
\eqn\pkea{ Q_2 = {1\over 2} \Tr (a_\phi^2)~,\quad Q_3 = {1\over 3}\Tr (a_\phi^3)~,\quad \ldots ~,\quad  Q_N = {1\over N} \Tr (a_\phi^N)~.}
Conjugate potentials $\sigma_n$ are identified in an analogous manner as in \mabb,
\eqn\pkeb{ a_t = \sigma_2 a_\phi + \sigma_3 \left(a_\phi^2-{I\over N}\Tr(a_\phi^3)\right) + \ldots + \sigma_N \left(a_\phi^{N-1}-{I\over N}\Tr(a_\phi^{N-1})\right)~. }
The idea is to work out the on-shell variation of the action, and then add terms such that we have $\delta I  \sim \sum_n Q_n \delta \sigma_n$.  The resulting action evaluated on-shell can then be identified with the black hole free energy in an ensemble in which the potentials $\{\sigma_n\}$ are held fixed. In \BanadosUE\ they find that a suitable action is
\eqn\pkec{  I = I_0 - \sum_{n=2}^N k(n-2)\sigma_n Q_n~,}
where $I_0$ is the original Chern-Simons action.   The on-shell variation of this action takes the desired form
\eqn\pked{ \delta I = 2k \sum_{n=2}^N Q_n\delta \sigma_n~,}
and its on-shell value is
\eqn\pkee{ I = -2k \sum_{n=2}^N (n-1) \sigma_n Q_n~.}
This gives a rather simple result for the free energy in the ensemble with fixed $\{\sigma_n\}$.   In this ensemble the charges $\{Q_n\}$
are functions of $\{\sigma_n\}$, obtained by solving the holonomy constraints.


\newsec{\hsl\ black holes}

Having constructed smooth black holes in $sl(3,R)$ gravity, we turn to the more challenging case of doing so in the three dimensional Vasiliev theory containing an infinite tower of higher spins $s\geq 2$. In addition to being interesting in its own right, it is this theory as opposed to the $sl(N,R)$ theories that is most likely to play some role in the possible relation of higher spin gravity to  string theory. The full Vasiliev theory has various ingredients which altogether describe the nonlinear coupling of the higher spin fields to matter, consistent with higher spin gauge invariance. We will not make use of most of them here: to make higher spin black holes, all we need from the theory is that its higher spin sector can be cast as two copies of Chern-Simons theory with connections valued in the infinite-dimensional Lie algebra \hsl. In particular, the matter fields of the Vasiliev theory are set to zero in the black hole background. For more details on the full theory, see \ProkushkinBQ, as well as recent summaries in \ChangMZ\ and \AmmonUA.

After describing \hsl\ in some detail, we will apply the logic of the previous chapter to a satisfying end, emerging with the asymptotic black hole partition function and hence a generalized Cardy formula for spin-3 charged black holes. Our results and their CFT counterparts \GaberdielYB\ lend support to the minimal model duality proposal.

In most of the following, we focus on the unbarred Chern-Simons connection with the barred sector implied. We work with the connection $a = a_zdz+a_{\zb}d\zb$ stripped of $\rho$-dependence as in \pkdj, and the $\rho$-dependence is restored by conjugation with a matrix we introduce momentarily.

\subsec{\hsl\ higher spin gravity}

The \hsl\ Lie algebra is spanned by generators labeled by two integers: a spin and a mode index. We use the notation of \GaberdielWB, in which a generator is represented as
\eqn\epa{V^s_m~, ~~ s \geq 2~, ~~ |m|<s~.}
The commutation relations are
\eqn\epb{[V^s_m,V^t_n] = \sum_{u=2,4,6,...}^{s+t-|s-t|-1}g^{st}_u(m,n;\l)V^{s+t-u}_{m+n}}
with structure constants given in Appendix A. The parameter $\l$ labels inequivalent algebras; while its appearance enters via generalized hypergeometric functions, these simplify to polynomials in $\l^2$ when evaluated for integer $(s,m)$.

The polynomial appearance of $\l^2$ is obvious upon constructing \hsl\ as a subspace of a quotient of the universal enveloping algebra of $sl(2,R)$ by its quadratic Casimir. For an $sl(2,R)$ with canonical commutation relation \pkdh, the Casimir $C_2$ is fixed as
\eqn\epc{C_2 = (L_0)^2-\half(L_1L_{-1}+L_{-1}L_1) = {1\over 4}(\l^2-1)~.}
The \hsl\ generators are then constructed from $sl(2,R)$ generators as\foot{\hsl\ also admits a symmetry under $V^s_m \rar (-1)^s V^s_m$, thus there is a second representation along the following lines.}
\eqn\epd{ V^s_m = (-1)^{m+s-1} {(m+s-1)!\over (2s-2)!}\underbrace{ \big[ L_{-1}, \ldots [L_{-1}, [L_{-1},}_{s-1-m} L_1^{s-1}]] \big]}
which generate \hsl\ upon modding out by the ideal formed by \epc\ and dropping the identity element, which is formally $V^1_0$.

The generators with $s=2$ form an $sl(2,R)$ subalgebra, and the remaining generators transform simply under the adjoint $sl(2,R)$ action as
\eqn\epe{[V^2_m,V^t_n] = (m(t-1)-n)V^t_{m+n}~.}
These $sl(2,R)$ generators will be relevant in the construction of the BTZ solution. In addition we can move from $a$ to $A$ connections a la \pkdj\ by conjugation with
\eqn\epg{b = e^{\rho V^2_0}}
giving generators with mode index $m$ a factor of $e^{m\rho}$.

In contrast to the $sl(N,R)$ algebras, \hsl\ has only a single $sl(2,R)$ subalgebra, nor does it have any other nontrivial subalgebra: any commutator of two generators each with spin $s>2$ will produce a generator with spin $t>s$. However, at $\l=N$ an ideal forms, consisting of all generators with spins $s>N$, and one recovers $sl(N,R)$ upon modding out by this ideal. This feature is manifest, for instance, in the simple low-spin commutator
\eqn\eph{[V^3_2,V^3_{-2}] = 8V^4_0 - {4\over 5}(\l^2-4)V^2_0}
upon setting $\l=2$. In this manner we can view the $sl(N,R)$ gravity theories as limiting cases of \hsl\ gravity\foot{This comes with the important caveat that at present, there is little understanding of how to couple matter to $sl(N,R)$ gauge fields, that is, how to truncate the full Vasiliev theory while still including the nontrivial fields besides the gauge fields.}.

When $\l=1/2$, this algebra is isomorphic to hs(1,1) whose commutator can be written as the antisymmetric part of the Moyal product. Similarly, the general $\l$ commutation relations \epb\ can be realized as a star commutator
\eqn\epj{[V^s_m,V^t_n]= V^s_m \star V^t_n - V^t_n \star V^s_m}
if we define the associative product
\eqn\epl{V^s_m \star V^t_n \equiv \half \sum_{u=1,2,3,...}^{s+t-|s-t|-1}g^{st}_u(m,n;\l)V^{s+t-u}_{m+n}~.}
This is known as the ``lone star product'' \PopeSR, and \epj\ follows upon using the fact that
\eqn\epm{\eqalign{g^{st}_u(m,n) &= (-1)^{u+1}g^{ts}_u(n,m)}~.}
In the remainder of this chapter, all multiplication of \hsl\ generators is done using this lone star product.

One can define a bilinear trace by picking out the $V^1_0$ element of any lone star product, up to some normalization:
\eqn\epn{\Tr (V^s_mV^t_n) \propto g^{st}_{s+t-1}(m,n;\l)\delta^{st}\delta_{m,-n}~.}
Explicitly, this structure constant can be written rather simply as
\eqn\epna{g^{ss}_{2s-1}(m,-m;\l) = (-1)^m{2^{3-2s}\Gamma(s+m)\Gamma(s-m)\over(2s-1)!!(2s-3)!!}\prod_{\sigma=1}^{s-1}(\l^2-\sigma^2)~.}
Note that the trace automatically factors out the ideal when $\l=N$: when we compute the holonomies of our black hole, this will allow easy comparison to the $sl(N,R)$ cases. To interface with the previous chapter's $sl(3,R)$ results and the conventions of \refs{\GutperleKF,\AmmonNK}, we choose to normalize our trace as
\eqn\epnb{\Tr (V^s_mV^s_{-m}) = {12\over (\l^2-1)}g^{ss}_{2s-1}(m,-m;\l)}
which implies the $s=2$ traces
\eqn\epnc{\Tr(V^2_1V^2_{-1})=-4~, ~~ \Tr(V^2_0V^2_{0})=2~.}

In \GaberdielWB, it was shown that the asymptotic symmetry group of \hsl\ gravity with generalized AdS boundary conditions \maaf\ is the infinite-dimensional $W$-algebra known as $W_{\infty}[\l]$.  To show this in the unbarred sector, say, one follows \pkdl\ by writing an \hsl-valued connection in highest-weight gauge, as
\eqn\epp{a_z = V^2_1 -{2\pi\over k}\ml(z)V^2_{-1} + \sum_{s=3}^{\infty}J^{(s)}(z)V^s_{1-s}}
where the $J^{(s)}(z)$ are spin-$s$ currents, and demanding that an infinitesimal gauge transformation leave the form of the connection intact. $W_{\infty}[\l]$ is a nonlinear algebra with OPEs now known in closed form \CampoleoniHG. It contains \hsl\ not as a proper subalgebra, but only in the infinite central charge limit in which the nonlinear terms drop out; from this perspective, \hsl\ is known as the ``wedge subalgebra'' of $W_{\infty}[\l]$. Expanding the currents in modes,
\eqn\epq{J^{(s)}(z) = \sum_{m\in Z}{J^{(s)}_m\over z^{m+s}}~,}
the \hsl\ generators are identified as the ``wedge modes''
\eqn\epr{J^{(s)}_m = V^s_m~, ~~ |m|<s~.}

To conclude this subsection, the black holes we will construct are dual to asymptotically high temperature states of generic CFTs with \winf\ symmetry, deformed by spin-3 chemical potential. Not only will this be a useful perspective as we compute in the bulk, but we will later discuss the dual CFT computation \GaberdielYB\ that yields perfect agreement with the bulk results.

\subsec{Building the \hsl\ black hole}

There are two main steps to constructing higher spin black holes in a generic theory of higher spin gravity with Lie algebra $\cg$ containing at least one $sl(2,R)$ subalgebra. These are preceded by a zeroth step, which is to write down the BTZ solution using generators of an $sl(2,R)$ subalgebra of $\cg$, and to compute its Euclidean time circle holonomy eigenvalues. The two steps are:\vs

{\it (i)} Write down the higher spin black hole connection with some nonzero higher spin chemical potentials.\vs

{\it (ii)} Make the black hole smooth. \vs

To find the right connection, we recall the $sl(3,R)$ black hole connection \maba\ in wormhole gauge: Ward identities showed that the leading term in $a_{\zb}$ represents a nonzero spin-3 chemical potential, and flatness fixed the rest of $a_{\zb}$ to be a simple traceless function of $a_z$. We are led to generalize this to the $\cg$ case: an unbarred connection for a higher spin black hole with nonzero spin-$s$ chemical potential $\mu_s$ can be written, in wormhole gauge, as
\eqn\ept{\eqalign{a_z &= a_z^{BTZ} + ({\rm higher~spin~charges})\cr
a_{\zb} &\sim \mu_s\left[(a_z)^{s-1}-{\rm trace}\right]\cr}}
where multiplication is determined by the chosen representation of the bulk Lie algebra, and the barred connection is as usual implied. To restore the $\rho$-dependence, one conjugates as in \pkdj\ by $b=e^{\rho L_0}$, where $L_0$ is the diagonal element of the $sl(2,R)$ subalgebra. Note that $a_z$ is simply the highest-weight gauge connection, as in \epp, with constant charges.

As a quick check, this will clearly give the correct asymptotic falloff at large $\rho$ for the chemical potential term: the leading piece of $(a_z)^{s-1}$ will carry $e^{(s-1)\rho}$ dependence. In the $\cg=$\hsl\ case, for instance, this is manifestly true because
\eqn\bjjj{(V^2_1)^{s-1} = V^{s}_{s-1}}
as is obvious from \epd.

As we have emphasized repeatedly in earlier sections, in order to make the black hole {\it smooth}, we must fix all charges in terms of the potentials $(\mu_s,\tau)$ consistent with an integrability condition \pkdv, where now $Q \sim J^{(s)}$ and $\alpha \sim \alpha_s\propto \mu_s$. To do so, we return to the BTZ holonomy condition: computing the Euclidean time circle holonomy matrix $\omega$ once again,
\eqn\epv{\omega = 2\pi( \t a_z+\bar{\t}a_{\zb})}
we demand that its eigenvalues equal those of the BTZ solution. One conveniently expresses this condition as
\eqn\epu{\Tr(\o^n)=\Tr(\o_{BTZ}^n)~,\quad n= 2, 3, \ldots, {\rm rk(\o)}~.}

Although the connections \ept\ are in wormhole gauge, the $sl(3,R)$ case encourages the perspective that somewhere on the gauge orbit of this solution lies a solution with a manifest black hole metric. Its smooth Euclidean horizon as seen by the tower of higher spin fields exactly reproduces the holonomy equations \epu. We assume this to be true in what follows.

Let us also comment that \epu\ works directly on the level of the matrix $\o$, that is, we do not exponentiate to determine the holonomy $H$. With regard to constructing smooth black hole solutions, this approach suffices to capture the essential feature of the BTZ black hole; in the case where the exponentiation of $\cg$ to a Lie group is not understood, neither is the notion of a trivial holonomy. This is the case with $\cg=~$\hsl. Nevertheless we continue to abuse language by referring to \epu\ as a holonomy condition.\foot{As \CastroIW\ showed for the case $\cg=sl(N,R)$, the space of solutions with trivial holonomy can include new solutions which are {\it not} black holes. It is this type of solution which relies on an understanding of the group manifold, and an interesting open challenge is to construct such conical defect-type solutions in the \hsl\ theory directly.}

Without further ado, let us apply this algorithm to construct the \hsl\ black hole with spin-3 chemical potential. First, the smooth BTZ solution is
\eqn\epua{\eqalign{a_z &= V^2_1 +{1\over 4\t^2}V^2_{-1}\cr
a_{\zb}&=0\cr}}
where we have used \pkbdb. The BTZ holonomy matrix $\o_{BTZ}$ is
\eqn\epub{\o_{BTZ} = 2\pi\t\left(V^2_1 +{1\over 4\t^2}V^2_{-1}\right)~.}
All odd-$n$ traces vanish. The lowest even-$n$ traces are
\eqn\epuc{\eqalign{\Tr(\omega_{BTZ}^2) &= -8\pi^2\cr
\Tr(\omega_{BTZ}^4) &= {8\pi^4\over 5}(3\l^2-7)\cr
\Tr(\omega_{BTZ}^6) &= -{8\pi^6\over 7}(3\l^4-18\l^2+31)~.\cr}}

The simplest higher spin black hole has spin-3 chemical potential, so following \ept\ our  ansatz is
\eqn\epud{\eqalign{a_z &= V^2_{1} -{2\pi\ml\over k}V^2_{-1} -N(\l){\pi\mw\over 2k}V^3_{-2} + J\cr
a_{\zb}&= -\mu N(\l)\left(a_z\star a_z - {2\pi\ml\over 3k}(\l^2-1)\right) \cr}}
where
\eqn\epue{J = J^{(4)}V^4_{-3}+J^{(5)}V^5_{-4}+\ldots}
allows for an infinite series of higher-spin charges.  The solution is accompanied by the analogous  barred connection.  $N(\l)$ is a normalization factor,
\eqn\epuf{N(\l) = \sqrt{20\over(\l^2-4)}}
chosen to simplify comparison to the $sl(3,R)$ results of the previous section. In particular, truncating all spins $s>3$ gives a solution with the same generator normalizations and bilinear traces as the spin-3 black hole of the \slth\ theory.

Suppressing the dependence on barred quantities, we think of this black hole as a saddle point contribution to the partition function
\eqn\epug{Z(\t,\a) = \Tr\left[e^{4\pi^2i(\t\ml+\a\mw)}\right]}
where we continue to define the potential as
\eqn\epuh{\a=\taub \mu}
where $\tau$ is the modular parameter of the boundary torus. This assignment will once again be justified upon solving the holonomy equations, as the charges will satisfy the integrability condition
\eqn\epui{{\p \Lc \over \p \alpha} = {\p \Wc \over \p \tau}~.}

It is instructive to compare the \hsl\ black hole to the $sl(3,R)$ black hole. There are three novel infinities here:\vs

\bul The number of holonomy equations \epu, by virtue of the infinite dimensionality of \hsl. Morally, we are fixing smoothness at the black hole horizon for the metric and an infinite tower of higher spin fields. \vs

\bul The UV behavior of the tower of metric-like higher spin fields. While there is no unambiguous prescription for how to write these fields in terms of traces over vielbeins beyond low spins \CampoleoniHG, the spin-$s$ field will generically involve traces over $s$ vielbeins. Recalling \pkde\ and our connection \epud\ (and its barred partner), metric-like higher spin fields grow with increasingly large powers of $e^{\rho}$ at successively higher spins. As with the metric, such behavior is gauge-dependent, and in contrast to the terms in the Chern-Simons connection it is not clear how to physically intepret these fields; but their rapid UV growth bears noting.\vs

\bul The number of nonzero higher spin charges $J^{(s)}$, as a generic solution of the infinite set of holonomy equations would dictate. This is particularly clear from a \winf\ perspective: thinking of taking operator expectation values
\eqn\epz{\langle J^{(s)}\rangle_{\a} = {\Tr\left[J^{(s)}e^{4\pi^2i(\t\ml+\a\mw)}\right]\over \Tr\left[e^{4\pi^2i(\t\ml+\a\mw)}\right]}}
and expanding perturbatively in $\a$, the coupled nature of the \winf\ OPEs guarantees that all charges are nonzero. In particular, one uncovers the following simple perturbative expansion,
\eqn\epza{\eqalign{\ml &= \ml_0+\a^2\ml_2+\ldots\cr
\mw&= \a\mw_1+\a^3\mw_3+\ldots\cr
J^{(s)} &= \a^{s-2}J^{(s)}_{s-2}+ \a^{s}J^{(s)}_{s}+\ldots~.}}
The polynomial degree and interdependence of the holonomy equations will force us to work perturbatively in $\a$.

\subsec{Black hole partition function}
The holonomy matrix $\o$ for the black hole connection \epud\ is
\eqn\epze{\o = 2\pi\Bigg[\tau a_z + \a N(\l)\left(a_z\star a_z - {2\pi\ml\over 3k}(\l^2-1)\right)\Bigg]~.}
Details on the mechanics of solving equations \epu, and the equations themselves, are given in \KrausDS. Suffice it to say here that this is a highly overconstrained problem which nevertheless has a solution. Using the perturbative expansions \epza, the charges through $O(\a^8)$ are given as
\eqn\epzf{\eqalign{\!\!\!\!\!\!\ml &= -{k\over 8\pi\t^2} + {5k\over 6\pi\t^6}\a^2-{50k\over3\pi\t^{10}}~{\l^2-7\over\l^2-4}
\a^4+{2600k\over 27\pi\t^{14}}~{5\l^4-85\l^2+377\over (\l^2-4)^2}\a^6\cr
&\quad  -{68000k\over 81\pi\t^{18}}~{20\l^6-600\l^4+6387\l^2-23357\over(\l^2-4)^3}
\a^8+\ldots\cr
\!\!\!\!\!\!\mw &= -{k\over 3\pi\t^5}\a+{200k\over27\pi\t^{9}}~{\l^2-7\over\l^2-4}\a^3
-{400k\over 9\pi\t^{13}}~{5\l^4-85\l^2+377\over (\l^2-4)^2}\a^5\cr
&\quad+{32000k\over 81\pi\t^{17}}~{20\l^6-600\l^4+6387\l^2-23357\over(\l^2-4)^3}
\a^7+\ldots\cr
\!\!\!\!\!\!J^{(4)} &= {35\over9\t^8}~{1\over\l^2-4}\a^2-{700\over9\t^{12}}~
{2\l^2-21\over(\l^2-4)^2} \a^4+{2800\over9\t^{16}}~{20\l^4-480\l^2+3189\over(\l^2-4)^3}
\a^6+\ldots\cr
\!\!\!\!\!\!J^{(5)} &= {100\sqrt{5}\over9\t^{11}}~{1\over(\l^2-4)^{3/2}}\a^3 -{400\sqrt{5}\over 27\t^{15}}~{44\l^2-635\over(\l^2-4)^{5/2}}\a^5+\ldots\cr
\!\!\!\!\!\!J^{(6)}&={14300\over 81\t^{14}}~{1\over (\l^2-4)^2}\a^4+\ldots~.\cr    }}
The charges $(\ml,\mw)$ manifestly satisfy the integrability condition; integrating either one of them, we recover the left-moving black hole partition function through $O(\a^8)$:
\eqn\epzg{\eqalign{ \ln Z(\t,\a) & =
{i \pi k \over 2\tau} \Bigg[1 -{ 4\over 3}~{\alpha^2 \over \tau^4}
+{400 \over 27}~ { \lambda^2-7 \over \lambda^2-4}~{\alpha^4\over \tau^8} -{1600 \over 27}~{5\lambda^4 -85 \lambda^2 +377 \over (\lambda^2-4)^2}~ {\alpha^6 \over \tau^{12}}\cr
&\quad\quad\quad~   +{32000\over 81}~{20\l^6-600\l^4+6387\l^2-23357\over(\l^2-4)^3}~{\a^8\over\t^{16}}\Bigg]+ \ldots~. }}
The leading term is the BTZ term \pkbh, using $\ln Z(\t,\a) = -I(\t,\a)$ and $c=6k$. This result reduces to the $sl(3,R)$ results of the previous section at $\l=3$, and has also been checked against independent calculations for $\l=4,5,\half$.

This is to be viewed as a Cardy formula for higher spin black holes with nonzero spin-3 chemical potential in CFTs with \winf\ symmetry.  It applies in the asymptotic regime
\eqn\epzh{\tau\rar 0~,~~ \a\rar 0~,~~ {\a\over\tau^2}~{\rm fixed}~.}
Given that we construct this solution in perturbation theory around the BTZ black hole, it is only sensible to think of this solution as a black hole too, now augmented by higher spin degrees of freedom contributing to its entropy. It is a delicate question whether this formula holds away from infinite temperatures, and for what CFTs. In the absence of a one-loop computation which has yet to be done, it is unclear when corrections to this result from perturbative excitations outside the black hole become important. If it holds for finite but large temperatures, any potential match to a CFT will involve more than its vacuum structure.

What is clear, however, is that at large enough temperatures, this result gives the higher spin charged black hole saddle point contribution to the Euclidean path integral. We now give more evidence for this from holography.

\subsec{Matching to CFT}

One might ask whether the partition function of the black hole in the $sl(3,R)$ theory --- which we computed exactly in the classical limit --- can be reproduced from some dual CFT calculation, but there are no known examples of unitary CFTs with $W_3$ symmetry and large central charge. The same goes for the series of $W_N$ minimal models at finite $N$.  But there {\it are} such examples of CFTs with  \winf\ symmetry.

A simple realization of $W_{\infty}[\l]$ symmetry at $\l=1$ is given by a theory of free, singlet complex bosons. The holomorphic and anti-holomorphic spin-$s$ currents are bilinear in the bosons, with $s$ derivatives peppered about. Accordingly, the symmetry algebra (also known \PopeSR\ as $W_{\infty}^{\rm PRS}$)  can be written in a linear basis, unlike the \winf\ algebra for all other values of $\l$. One can derive the exact partition function \epug\ in this theory at asymptotically high temperatures $\tau_2\rar 0$, which is
\eqn\epzi{\ln Z(\tau,\alpha) = -{3 ik \over 2\pi \tau} \int_0^\infty\! dx \left[ \ln \left(1-e^{-x + { ia \alpha  \over \tau^2}x^2 }\right) +
\ln \left(1-e^{-x - { ia \alpha  \over \tau^2}x^2 }\right)\right]}
where
\eqn\epzj{a = \sqrt{5\over 3\pi^2}~.}
Expanding in $\a$ gives the result \epzg\ at $\l=1$. We notice that \epzi\  is nonanalytic at $\alpha =0$, so its series expansion  has zero radius of convergence, as expected. A similar computation can be done at $\l=0$ with free fermions.

The reason these computations match the bulk result is due to symmetry. These free CFTs are not dual to Vasiliev gravity, if only on account of their $U(N)$ symmetry; to reproduce \epzg, apparently all that matters is that they furnish the asymptotic symmetry of the \hsl\ gravity theory. The partition function can be viewed in CFT as a perturbative sum of torus correlation functions. At asymptotically high temperatures, we can relate each term by a modular S transformation to the low temperature behavior, which is in turn determined by the vacuum symmetry algebra alone.

This is exactly the philosophy discussed in the introduction for computing the BTZ partition function, indeed it follows purely from the presence of 2d conformal symmetry, extended or not. For some more insight, we return to the other example discussed in section 2, the charged BTZ black hole. On the CFT side, the symmetry is Virasoro plus a U(1) current algebra; the left-moving partition function
\eqn\eja{Z(\t,z) = \Tr\left[e^{2\pi i\t L_0} e^{2\pi i z Q}\right]}
with U(1) charge $Q$ and conjugate potential $z$ transforms under a modular $S$-transformation as
\eqn\ejb{Z(\t'=-1/\t,z'=-z/\t) = e^{{\pi i c\over 3} {z^2\over \t}}Z(\t,z)}
where $c$ is the central charge \KrausNB. The entropy of the charged black hole quoted in \pkce\ follows entirely from \ejb\ and the on-shell action in global AdS with a flat U(1) connection, which can be computed from the action \pkca.

In the case of the higher spin black hole with spin-3 chemical potential, the left-moving \winf\ CFT partition function is
\eqn\ejc{Z(\t,\a) = \Tr\left[e^{2\pi i\t L_0} e^{2\pi i \a W_0}\right]}
with spin-3 charge $W_0$ and conjugate potential $\a$. Unlike the case of a spin-1 current insertion, we do not know the modular transformation property of this object. One can instead proceed perturbatively in $\a$, carrying out modular transformations term-by-term. This calculation was carried out in \GaberdielYB\ for arbitrary $\l$ through $O(\a^6)$, to perfect agreement with \epzg.

Shifting notation\foot{Note the shift $L_0 \rar L_0-{c\over 24}$ relative to the previous sections, as well as the identification $\mw={1\over 2\pi}W_0$ (as implied by integrability, cf. \pkdb).} to that of \GaberdielYB, we consider the partition function
\eqn\ejd{Z(\hat\tau, \alpha) = \Tr \Bigl( \hat{q}^{L_0 - {c\over 24}} \, y^{W_0} \Bigr)  \ , \qquad
\hat{q} = e^{2\pi i \hat{\tau}} \ , \quad y = e^{2\pi i \alpha} }
so that $\hat\t\rar 0$ is the high temperature limit. Expanding in $\a$,
\eqn\ejf{Z_(\hat\tau, \alpha) = \Tr \Bigl( \hat{q}^{L_0 - {c\over 24}} \Bigr) +
{(2\pi i \alpha)^2\over2!}  \Tr \Bigl( (W_0)^2\, \hat{q}^{L_0 - {c\over 24}} \Bigr) +
{(2\pi i \alpha)^4\over4!}  \Tr \Bigl( (W_0)^4\, \hat{q}^{L_0 - {c\over 24}} \Bigr) + \cdots ~.}
Our goal is to compute the modular transformation properties of these traces and evaluate in the vacuum. Thus we compute connected $2n$-point correlators of $W$ currents on the plane and map them to the infinite cylinder.

Let us present some of the relevant technology; the reader is referred to \GaberdielYB\ for the full scope of the computation, which includes new results on modular properties of traces with higher spin zero mode insertions for arbitrary spin. We specialize to the spin-3 Virasoro primary with conformal dimension 3. Generally, one can express traces of zero modes over some representation $r$ in terms of contour integrals of torus amplitudes $F_r$, which are function of the spin-3 current $W$ and $\t$, as follows:
\eqn\eje{\Tr_r\bigl (W_0)^n q^{L_0-{c\over 24}}\bigr = {1\over (2\pi i)^n}\prod_{j=1}^n \oint{dz_j\over z_j}F_r((W^1,z_1),\ldots,(W^n,z_n);\t)}
where $W_0$ is defined by the mode expansion
\eqn\ejf{W_0 = {1\over 2\pi i}\oint dz W(z)z^2~.}

The torus amplitude $F_r$ is doubly periodic under identifications $z_j\sim e^{2\pi i}z_j \sim q z_j$, and its modular transformation properties are known: in particular, under an $S$-transformation, we have \zhu\
\eqn\ejg{F_r((W^1,z_1),\ldots,(W^n,z_n);\hat{\t}=-1/\t) = \tau^{3n}\sum_{s}S_{rs}F_{s}((W^1,z_1^{\t}),\ldots,(W^n,z_n^{\t});\t)}
where $S_{rs}$ is the modular $S$-matrix. Therefore we can obtain the high temperature trace \eje\ via contour integrals of vacuum torus amplitudes. We are interested only in the partition function at asymptotically high temperatures, which means that we take all traces in the vacuum representation of \winf\ (i.e. $r=0$ in the above), ignoring subleading corrections. Consequently, we would like to simplify \ejg\ as much as we can in terms of traces over zero modes which annihilate the vacuum.

To actually evaluate these integrals, we want to exchange currents $W$ in the $F_r$ for zero modes $W_0$ and boil everything down to taking vacuum expectation values. We can do just this with the torus amplitude recursion relation
\eqn\ejh{\eqalign{&F((W^1,z_1),\ldots,(W^n,z_n);\t) = F(W_0;(W^2,z_2),\ldots,(W^n,z_n);\t) \cr&+\sum_{j=2}^n\sum_{m\in N_0} {\cal{P}}_{m+1}\bigl{z_j\over z_1},q\bigr F\bigl (W^2,z_2),\ldots,(W^1[m]W^j,z_j),\ldots,(W^n,z_n);\t\bigr}}
where the ${\cal{P}}_{m+1}$ are Weierstrass functions defined in \GaberdielYB, and the quantity $W[m]$ represents a linear combination of modes of $W$,
\eqn\ejha{W[m] = (2\pi i)^{-m-1}\sum_{j=m}^{\infty}c(3,j,m)W_{j-2}}
where the constants $c(3,j,m)$ are defined by the generating function
\eqn\ejhb{(\log(1+z))^m(1+z)^2 = \sum_{j= m}^{\infty}c(3,j,m)z^{j}~.}

In words, for every current in a torus amplitude, we can replace it by a zero mode as long as we add new terms encoding the interactions between currents. The first term on the right hand side of \ejh\ can be written as a trace over spin-3 zero modes and vertex operators $V(W^j,z_j)$,
\eqn\ejea{F(W_0;(W^2,z_2),\ldots,(W^n,z_n);\t) = \left(\prod_{j=2}^nz_j^3\right)\Tr\bigl W_0 V(W^1,z_1)\ldots V(W^n,z_n)q^{L_0-c/24}\bigr~.}
%
%

Further reducing these to traces over zero modes, the second term on the right hand side of \ejh\ can be reduced by repeated application of \ejh. Similarly, the first term can be  reduced with help of a recursion relation derived in \GaberdielYB\ that is the analog of \ejh, now for torus amplitudes with zero mode insertions. Using these recursion relations, one can eventually boil down the trace \eje\ to contour integrals of Weierstrass functions and their derivatives, multiplied by vacuum $n$-point correlators of $W$ currents. Precisely which correlators appear is determined by the series \ejha\ and the powers of $z$ required for nonzero Weierstrass integrals.

The preceding recursion relations and modular $S$-transformation properties generalize to insertions of Virasoro primary fields of arbitrary positive integral conformal dimension, and to arbitrary modular transformations \GaberdielYB.

An interesting aspect of this calculation is that the nonlinear parts of \winf\ OPEs are crucial for the success of the matching to the bulk partition function, despite the initial expectation that they are irrelevant in the large $c$ limit. These enter at $O(\a^6)$, where the CFT computation becomes rather hefty; this is in contrast to the bulk calculation using black hole holonomy which appears to encode the same information in a more efficient way. A direct understanding of how and ``why'' the bulk holonomy condition captures these features of the CFT would be desirable.

\newsec{Discussion}

We  close this review by presenting  a --- by no means complete ---  list of open problems and avenues for future research.

While one motivation for the study of higher spin theories is to  obtain solvable models without relying on supersymmetry, it would nevertheless be worthwhile to consider the supersymmetric generalizations of the higher spin gravities and the Gaberdiel-Gopakumar $AdS_3/CFT_2$ duality. In particular, it would be interesting to investigate the existence  and properties of BPS black holes in these theories; some work in this direction was recently initiated in \refs{\TanXI,\DattaKM}.

One of the main results of the work described in  this review is that  holonomy provides us with     the correct  criterion  for what constitutes  an admissible  smooth   geometry.  The holonomy conditions work for  the black hole  as well as the conical deficit/surplus geometries \CastroIW. It would be very interesting to find an independent derivation of the holonomy conditions from a more geometric point of view in  higher spin gravity.

For \hsl\ higher spin gravity the construction of black holes carrying higher spin charge utilizes a perturbation expansion in the higher spin charge $\alpha$.
As discussed in section 4 there have been impressive checks of this proposal
by relating the perturbative $\alpha$ expansion of the bulk  partition  function  to the corresponding result in the CFT.  However, the perturbation expansion produces a small charge perturbation around the uncharged BTZ black hole.
Whether the charged black hole solution can be generalized to finite values of $\alpha$ is an open problem, as  the perturbation series for the special values  $\lambda=0,1$ (and quite likely for general values of $\lambda$) is not convergent. Mathematically, the calculation of the holonomy of  a black hole with finite charge  would entail the construction of an exponential map  for the infinite dimensional algebra \hsl.  Whether such a map is possible to construct, i.e. whether its possible to construct  an analog of  the notion of a Lie group of \hsl,  is an open question.

The fact that matter can be consistently coupled to the \hsl\ higher spin gravity  provides us in principle with a  probe of the black hole geometry which is not gauge dependent. One can calculate scalar two-point correlators perturbatively in $\a$ in the black hole background of section 4, which should help clarify the extent to which this solution can rightfully be called a black hole. This is work in progress \PE; for now we note only that the pure gauge nature of the higher spin fields plays a crucial role in efficiently computing the correlator.

 The simplicity of the black hole solution in three dimensional higher spin gravity can be attributed to  the Chern-Simons like formulation of the higher spin dynamics.
In three dimensions all massless higher spin fields do not carry any propagating degrees of freedom and hence even the \hsl\ theory is almost topological (apart from matter  which can be consistently coupled). Furthermore the large amount of symmetry makes it feasible  to perform calculations of AdS/CFT correlators in the black hole background, which can help to address many basic questions of black hole physics, such as the formation and evaporation and the information paradox.

On the other hand, in  four dimensions it is quite complicated to construct nontrivial solutions of Vasiliev theory  for several reasons. First, the analog of the simple BTZ black hole solution is missing in four  dimensions. Second, the dynamics are considerably more complicated due to the fact that the higher spin fields are not topological anymore. Third, there is no known  analog of  the holonomy condition  described above, which diagnoses  the existence of a black hole in a gauge invariant manner.

Examples of  solutions in four dimensional Vasiliev theory  include \refs{\SezginRU,\SezginPV,\IazeollaWT,\IazeollaNF}, and those potentially related to black holes include  \refs{\DidenkoVA,\DidenkoTD,\IazeollaCB}. However the black hole character of the  solutions can only be seen  in a weak field approximation in the gravitational sector.   Recently  the finite temperature behavior  of vector models on the sphere in the large $N$ limit  was studied on the field theory side in \ShenkerZF.  It was found that the phase transition occurs at very high temperatures of order $\sqrt{N}$ and it was suggested that this behavior of the free energy  cannot  be described by a black hole in the bulk higher spin gravity as expected by the standard AdS/CFT philosophy.

An interesting question is whether it is possible to generalize the lessons learned from three dimensional higher spin gravity to higher dimensional higher spin theories. It may be the case that the advances in lower dimensions help us to construct the elusive  black hole solutions in four dimensional higher spin gravity.

\bigskip

\noindent{\bf Acknowledgments}

\medskip

This work was supported in part by NSF grant PHY-07-57702.

\appendix{A}{Some details on \hsl}
The \hsl\ structure constants are given as
\eqn\ejpa{g_u^{st}(m,n;\lambda) = {q^{u-2}\over2(u-1)!}\phi_u^{st}(\lambda)N_u^{st}(m,n)}
where
\eqn\ejpb{\eqalign{N_u^{st}(m,n) &= \sum_{k=0}^{u-1}(-1)^k
\left(\matrix{u-1 \cr k}\right)
[s-1+m]_{u-1-k}[s-1-m]_k[t-1+n]_k[t-1-n]_{u-1-k}\cr
\phi_u^{st}(\lambda) &= \ _4F_3\left[\matrix{\half + \lambda ~,~   \half - \lambda  ~,~ {2-u\over 2} ~ ,~ {1-u\over 2}\cr
{3\over 2}-s ~ , ~~ {3\over 2} -t~ ,~~ \half + s+t-u}\Bigg|1\right]\cr}}
We make use of the descending Pochhammer symbol,
\eqn\ejpc{[a]_n  = a(a-1)...(a-n+1)~.}
$q$ is a normalization constant that can be scaled away by taking $V^s_m \rar q^{s-2}V^s_m$. As in much of the existing literature, we choose to set $q=1/4$.

We note a handful of useful properties of the structure constants:
\eqn\ejpd{\eqalign{&\phi^{st}_u\left(\half\right)=\phi^{st}_2(\l)=1\cr
&N^{st}_u(m,n)=(-1)^{u+1}N^{ts}_u(n,m)\cr
&N^{st}_u(0,0)=0~, ~~ u ~{\rm even}\cr
&N^{st}_u(n,-n)=N^{ts}_u(n,-n)~.\cr}}
The first three of these imply, among other things, the isomorphism hs[$\half$]~$\cong$~hs(1,1); that the lone star product can be used to define the \hsl\ Lie algebra; and that all zero modes commute.

\listrefs
\end